\documentclass[a4paper,11pt]{article}
\pdfoutput=1 

\usepackage{jheppub} 

\usepackage[T1]{fontenc} 
\usepackage{float}
\usepackage{makecell}
\usepackage{tabstackengine}
\AtBeginEnvironment{pmatrix}{\setlength{\arraycolsep}{0.01pt}}
\newcommand{\ra}{\rightarrow}
\newcommand{\hpm}{H^\pm}
\newcommand{\hp}{H^\pm}
\newcommand{\mhp}{m_{H^\pm}}

\newcommand{\tb}[1][]{\ensuremath{\tan^{#1}\!\beta}}

\title{\boldmath New charged Higgs boson discovery channel at the LHC}

\author[a]{R. Benbrik,}
\author[a]{M. Boukidi,}
\author[b]{B. Manaut,}
\author[a]{M. Ouchemhou}
\author[c]{S. Semlali,}
\author[b]{and S. Taj}

\affiliation[a]{Polydisciplinary Faculty, Laboratory of Fundamental and Applied Physics, Cadi Ayyad University,\\Sidi Bouzid, B.P. 4162, Safi, Morocco.}
\affiliation[b]{Polydisciplinary Faculty, Laboratory of Research in Physics and Engineering Sciences,
	Team of Modern and Applied Physics, Sultan Moulay Slimane University, Beni Mellal 23000, Morocco.}
\affiliation[c]{School of Physics and Astronomy, University of Southampton, \\Southampton, SO17 1BJ, United Kingdom.}

\emailAdd{r.benbrik@uca.ma}
\emailAdd{mohammed.boukidi@ced.uca.ma}
\emailAdd{b.manaut@usms.ma}
\emailAdd{ouchemhou2@gmail.com}
\emailAdd{souad.semlali@soton.ac.uk}
\emailAdd{s.taj@usms.ma}
\abstract{The ATLAS and CMS experiments have an ambitious search program for charged Higgs bosons. The two main searches for $H^\pm$ at the LHC have traditionally been performed in the $\tau \nu$ and $t b$ decay channels, as they provide the opportunity to probe complementary regions of the Minimal SuperSymmetric Model (MSSM) parameter space. Charged Higgs bosons  may decay also to light quarks, $H^\pm \to cs/cb$, which represent an additional probe for the mass range below $m_t$. In this work, we focus on $H^\pm \to \mu \nu$ as an alternative channel in the context of two Higgs doublet model type III. We explored the prospect of looking  $pp\to tb \hp$, followed by $\hp\to\mu \nu$ signal at the LHC. Such a scenario appears in 2HDM type-III  where couplings of the charged Higgs are enhanced  to $\mu\nu$. Almost all the experimental searches rely 
	on the production and decay of the charged Higgs are taken into account. We show that for a such scenario, the above signal  is dominant for most of the parameter space, and $\hpm \to \mu\nu$ can be an excellent complementary search. Benchmarks points are proposed for further Monte Carlo analysis.}
\begin{document} 
\maketitle
\flushbottom
\section{Introduction}
The discovery of a Higgs boson by the ATLAS \cite{Aad:2012tfa}  and CMS \cite{Chatrchyan:2012ufa} collaborations in 2012, which behaves very closely to the predictions of the Standard Model (SM),  has played an extremely important role in the operation of the Large Hadron Colliders (LHC)  so far. In the other hand, the search for physics beyond the Standard Model (BSM) have  failed to yield any results, but there are interesting paths to investigate. Such Higgs boson was raised  from $SU(2)_L$ doublet field boosting, a viable interpretation of the  Higgs mechanism which lead to a spontaneous symmetry breaking of the electroweak symmetry. Several of these BSM theories argue for the presence of extra scalar particles. An eminent example is supersymmetry (SUSY). Introducing a new scalar boson for each SM fermion, resulting in a multiplicity of new particles. Extended models with simply a doublet scalar leads for example to  general two Higgs Doublet Model (2HDM) \cite{Gunion:1989we,Branco:2011iw}. 

Clearly, the most general implementation of 2HDM has non-diagonal fermionic couplings in the flavour space, and can therefore induce flavor-changing neutral current (FCNC) effects at the tree level, which might be not consistent with the actual observations.  The most basic alternative is to enforce a $Z_2$ symmetry, which prohibits unwanted non-diagonal terms.  Following the $Z_2$ charge assignments to scalars and fermions, this leads to four realizations of 2HDMs type-I, -II,-X and -Y. The differences among these four settings is the manner in which the Higgs bosons couples to fermions are classified. Type-I is defined where one doublet couples to all fermions. Type-II is the situation where one doublet couples to up quarks and the other to down quarks and leptons. Type-X is the model where one doublet couples to all quarks and the other to all leptons, while a Type-Y is build such that one doublet couples to up quarks and leptons and the other to down quarks. An alternative solution is to also assume the so-called Cheng-Sher ansatz within the fermion sector, forcing the non-diagonal Yukawa couplings to become relative to the mass of the relevant fermions, which is called type III 2HDMs \cite{Cheng:1987rs}.  In this scenario, the avoidance of tree-level FCNC is guaranteed automatically by considering the alignment in the flavor space of Yukawa matrices.   In the present study, the 2HDM type III will be thoroughly scrutinized.  The diagonal matrix elements are fixed in order to be consistent with $B$ physics. By assuming CP conservation option, and very minimal version of Higgs couplings, the 2HDM comes with at least 7 free parameters; those are: Higgs masses, $m_h$, $m_H$, $m_\hpm$, $m_A$; $\tan\beta=v_2/v_1$ which is the ratio of the vacuum expectation values of the two Higgs doublets fields; $\cos(\beta - \alpha)$ (or $\sin(\beta - \alpha)$) where $\alpha$ is the mixing angle of the CP-even Higgs states; and finally the $m^2_{12}$ free parameter. 

A powerful approach to probe extended Higgs sectors is to search directly for new particles in collision experiments.  In another way, extra Higgs bosons can be probed indirectly by precision measurements of the properties of the discovered Higgs boson, since the effects of extra Higgs bosons can be observed in the observable of the discovered Higgs boson by mixing Higgs states and/or loop corrections.  A clean signature of non-minimal Higgs sectors is the existence of charged Higgs bosons, as they are expected in extended Higgs models with multiple Higgs doublet fields. The LHC has performed extensive searches for the charged Higgs boson at Run-1 and Run-2 in practically every possible channel by both experiments ATLAS~\cite{Aad:2013hla,Aad:2014kga,ATLAS:2016avi,ATLAS:2021upq,ATLAS:2015nkq,ATLAS:2020jqj,ATLAS:2018gfm,ATLAS:2018ntn} and CMS~\cite{Khachatryan:2015qxa,Khachatryan:2015uua,CMS:2018dzl,Sirunyan:2019hkq} with different luminosity running the mass range from 80GeV to TeV scale. 

Both ATLAS~\cite{Aad:2014kga} and CMS~\cite{Sirunyan:2019hkq} experiments have set limits on ($m_{H^\pm}, \tan\beta$) plane from $H^\pm \to \tau\nu$ observations at $\sqrt{s}=8$ TeV and 13 TeV. Large $\tan\beta$ is excluded for light $m_{H^\pm}$. Moreover, intensive studies on light charged Higgs mass has been explored in 2HDM type-I, and type-X where the B physics constraints are weak. As at the LHC, the production cross-section $pp\to t \bar{t}$ followed by $t\to b H^+$ provides the main source of light charged Higgs boson for $\tau\nu$ channel together with  $cs$ and $cb$. Low values of $\tan\beta < 2.3(1.4)$ are excluded for light charged Higgs mass up to top quark mass for type-III and type-I, respectively. The type-III  2HDM at intermediate $\tb$ is one of the  interesting discovery potential. In such scenario, the typical production of charged Higgs via the top channel becomes similar as type-I, but with very different behavior on charged Higgs branching ratio, resulting in spectacular signatures.

We have identified the potential importance of the $\hpm$ bosonic decay model in various analyses. Associated charged Higgs production~\cite{Coleppa:2014cca,Enberg:2014pua,Kling:2015uba,Akeroyd:2016ymd,Arhrib:2016wpw,Alves:2017snd,Arhrib:2019ywg,Arhrib:2020tqk}. Alternatively, the electroweak production of $\hpm$~\cite{Kanemura:2001hz,Cao:2003tr,Belyaev:2006rf,Bahl:2021str,Ferrari:2021hqc} in partnership with a neutral (pseudo)scalar depends on gauge coupling and dominates over gauge coupling. For type-I 2HDM, the electroweak production of a charged light Higgs can result in a multi-photon or multi-boson final state~\cite{Wang:2021pxc,Arhrib:2017wmo,Enberg:2018pye,Enberg:2018nfv,Arhrib:2021xmc}.

The purpose of this paper is to explore $pp \to t b H^\pm, H^\pm \to \mu \nu $ in the current context of LHC, to assess the extent to which they might complement the searches for light charged Higgs.  We will demonstrate that the production rates of such alternative production channel, in type-III, have the potential to be overwhelmingly stronger than the production channels followed $H^\pm \to \tau \nu$ .  More specifically, we will illustrate that the leptonic decay of a lightly charged Higgs boson, could be dominant and give alternative reachable signatures to those arising from top quark production and decay.  Consequently, such modes can serve as new discovery channels for light $ H^\pm$ states at the LHC. We show that the proposed signal will discover or rule out a significant  parameter space for Higgs scalars. The proposed signal complement the existing search strategies to expand the reach of LHC searches for charged Higgs. 

The paper is organized as follows: In Sec.~2, we briefly discuss the type-III 2HDM model, 
including the Yukawa interaction. In Sec.3 we discuss the existing theoretical and experimental bounds on the parameter space. Then we motivate towards 
$pp\to t b H^\pm, H^\pm \to \mu\nu$ signature in Sec.~4 where we also provide the details of the numerical analysis and we conclude in Sec.5.

\section{General 2HDM}
The general 2HDM is one of the minimal extension of the SM Higgs sector, which consist of two doublet scalar fields $\Phi_i$ (i =1, 2) with hypercharge $Y = +1$. The most general 2HDM potential is given by \cite{Branco:2011iw}:
\begin{eqnarray}
	V_{\rm{Higgs}}(\Phi_1,\Phi_2) &=& \lambda_1(\Phi_1^\dagger\Phi_1)^2 +
	\lambda_2(\Phi_2^\dagger\Phi_2)^2 +
	\lambda_3(\Phi_1^\dagger\Phi_1)(\Phi_2^\dagger\Phi_2) +
	\lambda_4(\Phi_1^\dagger\Phi_2)(\Phi_2^\dagger\Phi_1) \nonumber  +~\nonumber\\ && +
	\frac12\left[\lambda_5(\Phi_1^\dagger\Phi_2)^2 +\rm{h.c.}\right]
	+~\left\{\left[\lambda_6(\Phi_1^\dagger\Phi_1)+\lambda_7(\Phi_2^\dagger\Phi_2)\right]
	(\Phi_1^\dagger\Phi_2)+\rm{h.c.}\right\} \nonumber \\ && 
	-~\left\{m_{11}^2 \Phi_1^\dagger \Phi_1+ m_{22}^2\Phi_2^\dagger
	\Phi_2 + \left[m_{12}^2
	\Phi_1^\dagger \Phi_2 + \rm{h.c.}\right] \right\}\,.\label{CTHDMpot}
\end{eqnarray}

Following the hermiticity of Eq. (\ref{CTHDMpot}),$m_{11}^2$, $m_{22}^2$ and $\lambda_{1,2,3,4}$ are real parameters, whereas $\lambda_{5,6,7}$ and $m_{12}^2$ can be complex. Assuming the CP conservation, $\lambda_{5}$, $m_{12}^2$ are real. In the models where $Z_2$ symmetry is extended to the Yukawa sector to suppress the flavors changing neutral currents (FCNC) at tree level, terms that are proportional to $\lambda_6$ and $\lambda_7$ in the scalar potential are absent. 
Note that in 2HDM Type-III with the off diagonal elements in Yukawa matrices\cite{Cheng:1987rs, Diaz-Cruz:2004wsi}, $\lambda_7$ and $\lambda_6$ could be kept. In order to simplify the numerical analysis and to reduce the scanned parameters, it is more reasonable in this study to assume that in type-III, $\lambda_6 = \lambda_7 = 0$. After the electroweak symmetry breaking, we left with only 7 independent parameters: $m_{H^{\pm}}, m_{A}, m_{H}, m_{h}, \cos(\alpha-\beta), \tan \beta\ \text{and} \ m_{12}^2$.
	
\subsection{Yukawa interaction}
We adopt the description presented in \cite{Cheng:1987rs, Diaz-Cruz:2004wsi} to keep the FCNC under control, while inducing flavor violating Higgs signals, by assuming a flavor symmetry that suggest a specific texture of the Yukawa matrices, where the non-diagonal Yukawa couplings, $\tilde{Y}_{ij}$, are given in terms of fermions masses and dimensionless real parameter, $\tilde{Y}_{ij} \propto \sqrt{m_i m_j}/ v ~\chi_{ij}$. 

After the electroweak symmetry breaking(EWSB), we can rewrite the Yukawa Lagrangian in terms of physical scalar masses as:
\begin{eqnarray}
	-{\cal L}^{III}_Y &=& \sum_{f=u,d,\ell} \frac{m^f_j }{v} \left[ (\xi^f_h)_{ij}  \bar f_{Li}  f_{Rj}  h + (\xi^f_H)_{ij} \bar f_{Li}  f_{Rj} H - i (\xi^f_A)_{ij} \bar f_{Li}  f_{Rj} A \right] \nonumber \\
	&+& \frac{\sqrt{2}}{v} \sum_{k=1}^3 \bar u_{i} \left[ \left( m^u_i  (\xi^{u*}_A)_{ki}  V_{kj} P_L + V_{ik}  (\xi^d_A)_{kj}  m^d_j P_R \right) \right] d_{j}  H^+  \nonumber \\
	&+& \frac{\sqrt{2}}{v}  \bar \nu_i  (\xi^\ell_A)_{ij} m^\ell_j P_R \ell_j H^+ + H.c.\,, \label{eq:Yukawa_CH}
\end{eqnarray} 
In Table~\ref{coupIII}, we describe the Yukawa interactions in type-III in terms of the mixing angles and the free parameters $\chi_{ij}^f$. 

\begin{table}[H]
	\begin{center}
		\setlength{\tabcolsep}{8pt}
		\begin{tabular}{c|c|c|c} \hline\hline 
			$\phi$  & $(\xi^u_{\phi})_{ij}$ &  $(\xi^d_{\phi})_{ij}$ &  $(\xi^\ell_{\phi})_{ij}$  \\   \hline
			$h$~ 
			& ~ $  \frac{c_\alpha}{s_\beta} \delta_{ij} -  \frac{c_{\beta-\alpha}}{\sqrt{2}s_\beta}  \sqrt{\frac{m^u_i}{m^u_j}} \chi^u_{ij}$~
			& ~ $ -\frac{s_\alpha}{c_\beta} \delta_{ij} +  \frac{c_{\beta-\alpha}}{\sqrt{2}c_\beta} \sqrt{\frac{m^d_i}{m^d_j}}\chi^d_{ij}$~
			& ~ $ -\frac{s_\alpha}{c_\beta} \delta_{ij} + \frac{c_{\beta-\alpha}}{\sqrt{2}c_\beta} \sqrt{\frac{m^\ell_i}{m^\ell_j}}  \chi^\ell_{ij}$ ~ \\
			$H$~
			& $ \frac{s_\alpha}{s_\beta} \delta_{ij} + \frac{s_{\beta-\alpha}}{\sqrt{2}s_\beta} \sqrt{\frac{m^u_i}{m^u_j}} \chi^u_{ij} $
			& $ \frac{c_\alpha}{c_\beta} \delta_{ij} - \frac{s_{\beta-\alpha}}{\sqrt{2}c_\beta} \sqrt{\frac{m^d_i}{m^d_j}}\chi^d_{ij} $ 
			& $ \frac{c_\alpha}{c_\beta} \delta_{ij} -  \frac{s_{\beta-\alpha}}{\sqrt{2}c_\beta} \sqrt{\frac{m^\ell_i}{m^\ell_j}}  \chi^\ell_{ij}$ \\
			$A$~  
			& $ \frac{1}{t_\beta} \delta_{ij}- \frac{1}{\sqrt{2}s_\beta} \sqrt{\frac{m^u_i}{m^u_j}} \chi^u_{ij} $  
			& $ t_\beta \delta_{ij} - \frac{1}{\sqrt{2}c_\beta} \sqrt{\frac{m^d_i}{m^d_j}}\chi^d_{ij}$  
			& $t_\beta \delta_{ij} -  \frac{1}{\sqrt{2}c_\beta} \sqrt{\frac{m^\ell_i}{m^\ell_j}}  \chi^\ell_{ij}$ \\ \hline \hline 
		\end{tabular}
	\end{center}
	\caption {Yukawa couplings of the $h$, $H$, and $A$ bosons to the quarks and leptons in type-III 2HDM} 
	\label{coupIII}
\end{table}
Note that the off diagonal terms of Yukawa matrices can be present, if $i \neq j$. However, in our study, we will keep the version of 2HDM type-III without off diagonal terms in the Yukawa texture as shown in Eq~\ref{eq:Yukawa_CH} where $\chi_{ij}^f = 0$, for $i \neq j$.
The values of the on-diagonal elements in the Yukawa texture are as follows:
\setstacktabbedgap{5pt}
\begin{equation}
	\chi^u=\begin{pmatrix}
		0.187 & 0 & 0 \\
		0 & 0.254 & 0\\
		0 & 0 & 0.210
	\end{pmatrix},
	\chi^d=\begin{pmatrix}
		-0.553 & 0 & 0 \\
		0 & 2.863 & 0\\
		0 & 0 & 1.440
	\end{pmatrix},
	\chi^l=\begin{pmatrix}
		0.484 & 0 & 0 \\
		0 & -2.101 & 0\\
		0 & 0 & 1.400
	\end{pmatrix}
	\label{chi_bestIII}
\end{equation}
\textcolor{black}{The free parameters $\chi_{ij}^{u,d,l}$ are tested at the current constraints from B-physics observables by using \texttt{SuperIso}}~\cite{superIso}.
\section{Theoretical and experimental bounds}
In order to represent a viable BSM, the 2HDM has to respect certain theoretical and experimental constraints.
The theoretical requirements included in 2HDM are perturbativity of the scalar quartic couplings, vacuum stability, and the tree-level perturbative unitarity conditions for various scattering amplitudes of gauge boson and Higgs state, these constraints reflect some important conditions in the space parameter of the model which will exhibit in the following:

\begin{itemize}
	\item Unitarity constraints impose to a variety of scattering process to be unitary;scalar-scalar,gauge-boson–gauge boson, and scalar-gauge-boson, such  conditions implies a set of inequality that have to be fulfilled on the eigenvalues of scattering matrix as follows $|e_i|< 8\pi$ ~\cite{uni1,uni2,uni3}
	
	\item Perturbativity constraints impose to the quartic couplings of the scalar potential to obey the following conditions:$|\lambda_i|<8\pi$ ~\cite{Branco:2011iw}.
	
	\item Vacuum stability constraints require the potential to be bounded from below and positive in any direction of the field $\Phi_i$ , consequently, the space parameter must release the following condition~\cite{Barroso:2013awa,sta}:
	\begin{align}
		\lambda_1 > 0,\quad\lambda_2>0, \quad\lambda_3>-\sqrt{\lambda_1\lambda_2} ,\quad \lambda_3+\lambda_4-|\lambda_5|>-\sqrt{\lambda_1\lambda_2}
	\end{align}
	Besides the aforementioned theoretical constraints, we consider the experimental constraints comes from Electroweak Precision Observables (EWPOs),the null-searches from LHC,LEP and Tevatron experiments, furthermore we requires agreement with B-physics observable:
	\item Electroweak precision tests (EWPT):the agreement with electroweak precision observables(EWPOs), parametrized through the electroweak oblique parameters S, T, U at $95\%$ C.L are required, their experimental limit are set as following ~\cite{Grimus:2007if,oblique2,Haller:2018nnx}.
		
	\begin{align}
		S = 0.05 \pm 0.11,\quad T = 0.09 \pm 0.13,\quad U = 0.01 \pm 0.11.
	\end{align}
	\item Collider constraints: we ask for agreement with the limits obtained from various searches of additional Higgs bosons at the LHC as well as the requirement that one of the CP-even Higgs Boson should match the properties of the observed SM-like Higgs boson.We evaluate the former constraints with the tools \textbf{HiggsBouns-5.9.0}\cite{HB, Bechtle:2015pma} and the latter with the tools \textbf{HiggsSignal-2.6.0}\cite{HS}.
	
	\item Flavour constraints: In our analysis we have used the B-physics observable cite on the Table.2, where we use the tools \textbf{Superiso v4.1}\cite{superIso} to calculate them.
\end{itemize}
The B physics observable that we have took are as following:
{\renewcommand{\arraystretch}{1.5} 
	{\setlength{\tabcolsep}{0.1cm} 
		\begin{table}[H]
			\centering
			\setlength{\tabcolsep}{7pt}
			\renewcommand{\arraystretch}{1.2} %
			\begin{tabular}{|l||c|c|}
				\hline
				Observable&Experimental result&95\% C.L. Bounds\\\hline
				BR($B_{\mu}\to \tau\nu$)\cite{Haller:2018nnx}&$(1.06 \pm 0.19) \times 10^{-4}$&$ [0.68\times 10^{-4} , 1.44\times 10^{-4} ]$\\\hline
				BR($B_{s}^{0}\to \mu^{+}\mu^{-}$)\cite{Haller:2018nnx}&$(2.8 \pm 0.7) \times 10^{-9}$&$[1.4 \times 10^{-9}, 4.2\times 10^{-9}]$\\\hline
				BR($B_{d}^{0}\to \mu^{+}\mu^{-}$)\cite{superIso}&$(3.9\pm 1.5)\times10^{-10}$&$[0.9\times 10^{-10}, 6.9\times10^{-9}$\\\hline
				BR($\bar{B}\to X_{s}\gamma$)\cite{Bphys1,Haller:2018nnx}&$(3.32\pm 0.15)\times10^{-4}$&$[3.02\times 10^{-4} , 3.61\times 10^{-4}]$\\\hline
			\end{tabular}
			\caption{Experimental results of flavor observables: $B_{\mu}\to \tau\nu$, $B_{s,d}^{0}\to \mu^{+}\mu^{-}$ and $\bar{B}\to X_{s}\gamma$ at 95$\%$ C.L. Bounds.}
			\label{Tab2}
		\end{table}
\subsection{Cross sections and branching ratios}
		We use the public tools 2HDMC-1.8\cite{2HDMC} for checking the aforementioned theoretical constraint, the electroweak precision observables(EWPOs) as well as the different physical quantities pertinent to this phenomenological study.

		We point out some of the recent constraints from searches for the charged Higgs boson. The ATLAS collaboration has set an upper limit between 0.25\% and 0.031\% for the branching fraction $B(t \ra b \hpm)\times B(\hpm \ra \tau^\pm \nu_{\tau})$ in the range of masses varying from 90 GeV to 160 GeV after assuming the Standard Model cross-section for $t\overline{t}$ production~\cite{Aaboud:2018gjj}. An other search for charged Higgs boson in the same decay channel ($\hpm \ra \tau^\pm \nu_{\tau}$) for $\hpm$ in the mass range of 80 GeV to 3 TeV is presented by ATLAS group~\cite{Sirunyan:2019hkq}.
		In addition, the CMS collaboration has recently searched for a light charged Higgs boson with  mass in the range 100 GeV to 160 GeV and a CP-odd Higgs boson with $15~\text{GeV}<m_A< 75~~\text{GeV}$ in the decay chain, $t \ra b\hpm \ra bW^\pm A \to b W^\pm \mu^+ \mu^-$, although, no significant excess is observed, upper limits are set on the product of branching fractions, $B(t \ra b\hpm) \times B(\hpm \ra W^+ A) \times B(A\ra \mu^+ \mu^-)$ depending on $\mhp$ and $m_A$~\cite{CMS:2019idx}.
		Furthermore, the ATLAS collaboration has also lately presented several searches for a heavier charged Higgs decaying into a top and bottom quarks, for $\mhp > m_t +m_b$. No significant excess is reported and upper limits are set on $\sigma(\hp)\times B(\hp \ra t b)$ as a function of $\mhp$~\cite{CMS:2019rlz,ATLAS:2018ntn,CMS:2020imj}.

\subsection{B physics constraints}
As a first check, on the B physics constraints using the public package SuperIso \cite{superIso}, we present in Figure \ref{figbb}, the relevant constraints related to flavor observables: $B_{\mu}\to \tau\nu$, $B_{s,d}^{0}\to \mu^{+}\mu^{-}$ and $\bar{B}\to X_{s}\gamma$ which compatible the measurements at 95$\%$ C.L in both type-I(left) and type-III (right) panels. As for experimental limit we adopt LHCb collaboration \cite{LHCb:2012skj, LHCb:2017rmj} and from PDG book \cite{Tanabashi}. The $B_{s,d}^{0}\to \mu^{+}\mu^{-}$ decay mode receive contributions from CP-odd Higgs boson proportional to $(\xi^\ell_{A})^6/m^4_A$ \cite{Babu:1999hn}. As we can see, the possibility for light charged Higgs together with small $\tan\beta$ still compatible with recent measurements. From Fig \ref{figbb} one can read  that  small values of $\tan\beta$ ($<1.3$) are excluded by flavor physics constraints for different masses of $H^\pm$ in type-III, unlike type-I.
\begin{figure}
	\centering
	\includegraphics[height=8.cm,width=14.5cm]{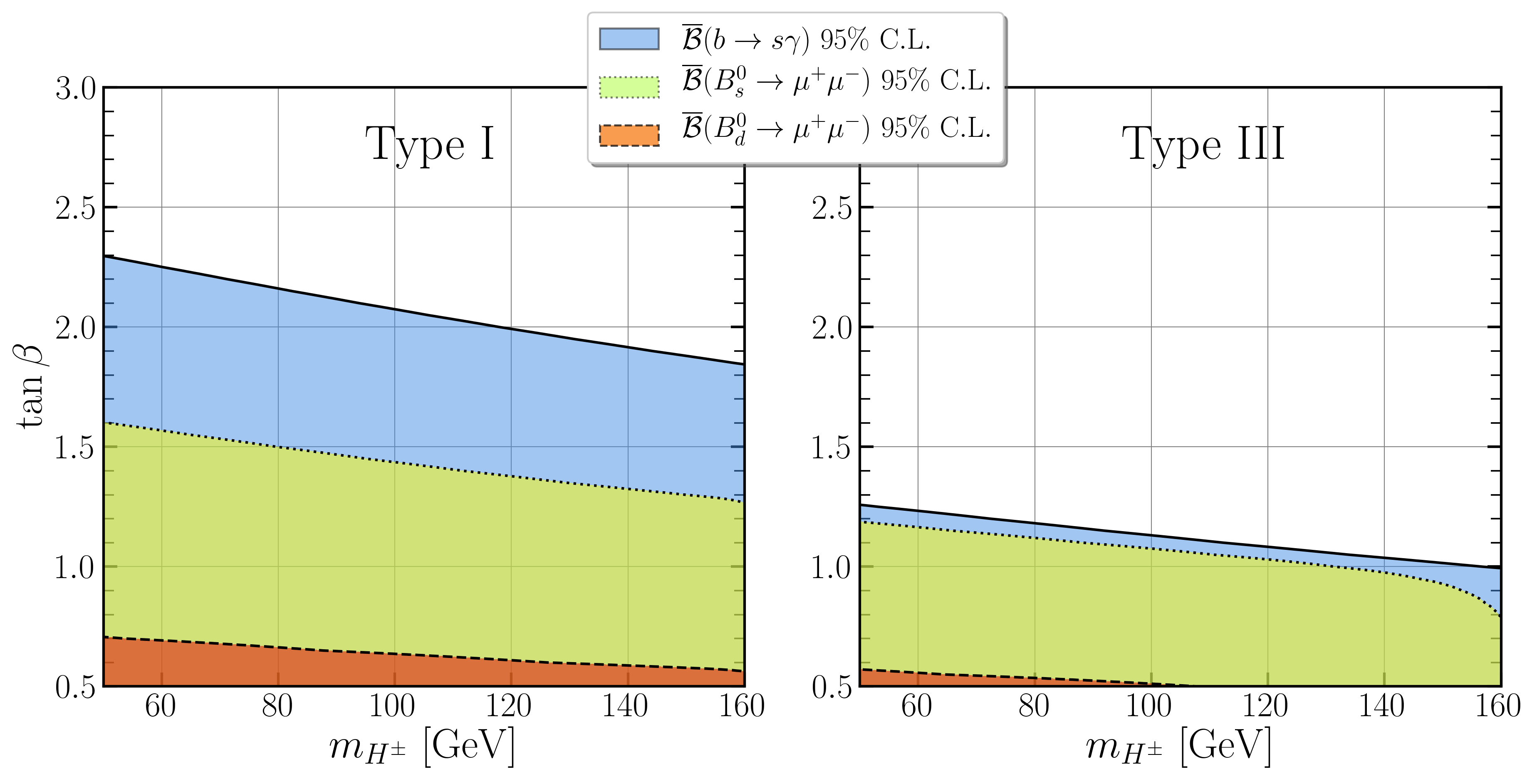}
	\caption{Excluded regions of the $(m_{H^\pm},~\tan\beta)$ parameter space  by flavor constraints at $95\%$ C.L.}
	\label{figbb}
	
\end{figure}
		\section{Numerical results}	
		The extended scalar sector of CP-conserving 2HDM can allow two interesting possibilities which have been the subject of many studies, where one of the two CP-even Higgses can be identified as the observed 126 GeV state at the LHC, whose properties are approaching those of the SM. For comprehensive reviews, see~\cite{Bernon:2015wef,Bernon:2015qea}. In the first scenario, the lightest Higgs is the observed Higgs particle at LHC ($h \equiv H_{SM}$). This can be achieved with or without decoupling of the other scalars. In this case, the Higgs couplings approach towards those of the standard model when $\cos(\beta -\alpha) \ra 0$, whereas in the second scenario, the heavier CP-even state ($H \equiv H_{SM}$) is identified as the SM-like Higgs whose couplings to gauge bosons approach the SM limit when $\sin(\beta -\alpha) \ra 0$. The focus of this study is nearly alignment limit without decoupling. This choice is well motivated by the interesting phenomenology of the light charged and/or neutral Higgs in 2HDM, particularly in type-III. We promote here a new signature to search for charged Higgs boson, while taking advantage of the Yukawa-texture contribution in $H^\pm$ decay width. For that, we employ all the recent results from the searches of light scalars and the signal measurements from LHC at Run-1 and Run-2, while taking advantage of the significant increase of Luminosity and energy, which began with a collision energy of 8 TeV rising to 13 TeV.
		
		The light charged Higgs is searched for in top quark decay channel, $t \ra \hp b$, for $\mhp< m_t-m_b$. The LHC experiments are looking for such a particle in the leptonic and quark-Charged Higgs decay channels, namely $ \tau \nu $, $ c \overline{s} $ and $ t^*\overline{b} $. 
		We recall that the previous searches in the decay channel $pp \ra t\hp \ra \tau^\pm \nu_{\tau}$ at Run-1, lead to upper limits on $B(t \ra b\hp)\times B(\hp \ra \tau \nu_{\tau})$ in the mass range $\mhp = 80-160~\text{GeV}$~~\cite{Aad:2014kga, Khachatryan:2015qxa,CMS:2016szv}. Moreover, the CMS collaboration has also recently set an upper limit (2\%~-~3\%) on $Br(t \ra \hp b)$ for $\text{80}~\text{GeV}<\mhp<160~\text{GeV}$, while assuming $Br(\hp \ra \tau \nu) =100\%$~\cite{Tanabashi}. In addition, ATLAS and CMS have set an upper limit of 20\% on $Br(t \ra \hp b)$ for the case $Br(\hp \ra  c \overline{s}) =100\%$ in the mass range of 90~GeV to 160 GeV~\cite{Tanabashi}. 
	\begin{table}[H]
	\centering
	\renewcommand{\arraystretch}{1.2} %
	\setlength{\tabcolsep}{0.5pt}
	\begin{tabular}{|c|c|c|c|c|c|c|c|}\hline
		&$m_h~[\mathrm{GeV}]$&$m_H~[\mathrm{GeV}]$&$m_A~[\mathrm{GeV}]$&	$m_{H^\pm}~[\mathrm{GeV}]$& $s_{\beta-\alpha}$&$\tan\beta$&$m_{12}^2 $ \\\hline
		$h$ Scenario	&$125$&$135$&$220$&$[50;\,160]$& $-0.98$&$[0.5;\,15]$&$m_h^2\tan\beta/(1+{\tan^{2}\beta})$\\\hline
		$H$ Scenario	&$95$&$125$&$177$&$[50;\,160]$& $-0.05$&$[0.5;\,15]$&$m_h^2\tan\beta/(1+{\tan^{2}\beta})$\\\hline
	\end{tabular}		
	\caption{2HDM input parameters for both scenarios.} \label{2hdm_par}
\end{table}
\begin{table}[H]
	\centering
	\renewcommand{\arraystretch}{1} %
	\setlength{\tabcolsep}{35pt}
	\begin{tabular}{|c|c|} \hline
		Abriviation& Expression    \\ \hline
		$\sigma_{2t}^{H\pm}(\bar{t}+b+\mu\nu)$ & $2\times\sigma(pp\to t\bar{t}\to H^{\pm}b+\bar{t}\to \mu\nu + b\bar{t})$  \\ \hline
		$\sigma_{2t}^{H\pm}(\bar{t}+b+\tau\nu)$ & $2\times\sigma(pp\to t\bar{t}\to H^{\pm}b+\bar{t}\to \tau\nu + b\bar{t})$  \\ \hline
		$\sigma_{2t}^{H\pm}(\bar{t}+b+c\bar{s})$ & $2\times\sigma(pp\to t\bar{t}\to H^{\pm}b+\bar{t}\to c\bar{s} + b\bar{t})$  \\ \hline	
		$\sigma_{2t}^{H\pm}(\bar{t}+b+t^*\bar{b})$ & $2\times\sigma(pp\to t\bar{t}\to H^{\pm}b+\bar{t}\to t^*\bar{b} + b\bar{t})$  \\ \hline
		$\sigma_{2t}^{H\pm}(\bar{t}+b+W^(*)h)$ & $2\times\sigma(pp\to t\bar{t}\to H^{\pm}b+\bar{t}\to W^(*)h + b\bar{t})$  \\ \hline 
	\end{tabular}
	\caption{The signal of charged Higgs production and its decays.} 
\end{table}	

\subsection{Part-I: $h$ Scenario}
In this scenario, we assume that the Higgs-like particle is $h$ with $m_h=125$ GeV. We perform a systematic scan over the 2HDM parameters as indicated in Table \ref{2hdm_par}:
After performing a scan over the 2HDM parameters, we show in Figure 2, the allowed region of parameter space in $(m_{H^\pm} , \tan \beta)$ plane by the theoretical and the current experimental bounds. The yellow and green regions are respectively excluded by LEP \cite{Abbiendi:2013hk} and LHC searches (see Fig \ref{fig12}(left panel) in the Appendix). The dotted line indicates compatibility with the observed Higgs signal at $2\sigma$. The majority of the region is excluded by theoretical and experimental constrains as can be seen from Table \ref{constraints_percent}, which demonstrates that only 4.31\% of the parameter space is permitted. On the other hand, one can read from the same figure that small values of $\tan\beta$ are allowed for different masses of $H^\pm$  in type-III, unlike type-I which excludes small $\tan\beta$  for
light $m_{H^\pm}$. Type-I is scrutinized by charged Higgs searches $H^\pm \to \tau\nu$  channel and by several flavour observables such as $\delta m_s$ and $B_{d,s} \to\mu^+\mu^-$ ~\cite{Sirunyan:2018hoz}.

\begin{figure}[H]
	\centering
	\begin{minipage}{0.51\textwidth}
		\centering
		\includegraphics[height=6.cm,width=7.0cm]{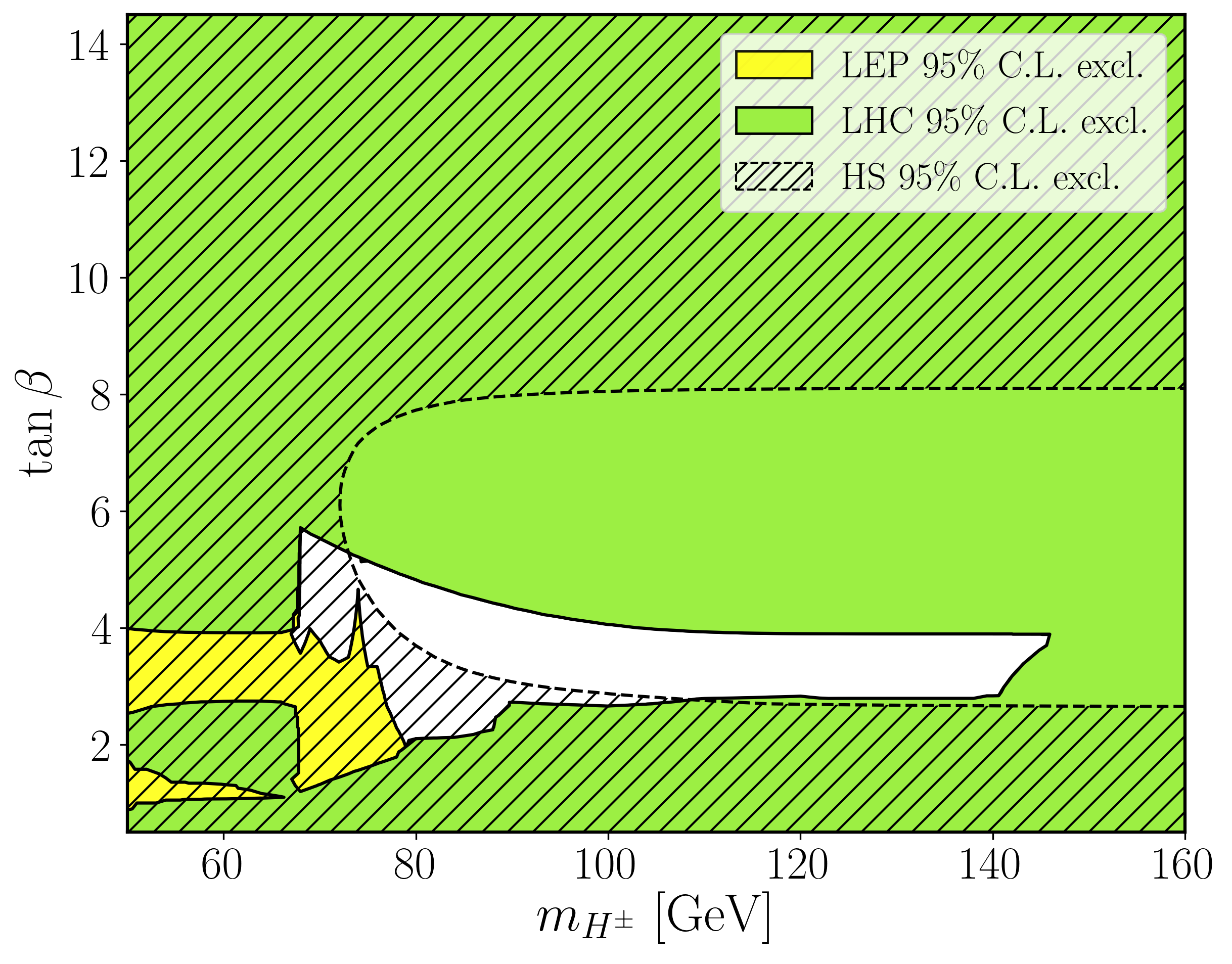}
	\end{minipage}
	\begin{minipage}{0.45\textwidth}
		\centering
		\includegraphics[height=6.cm,width=7.0cm]{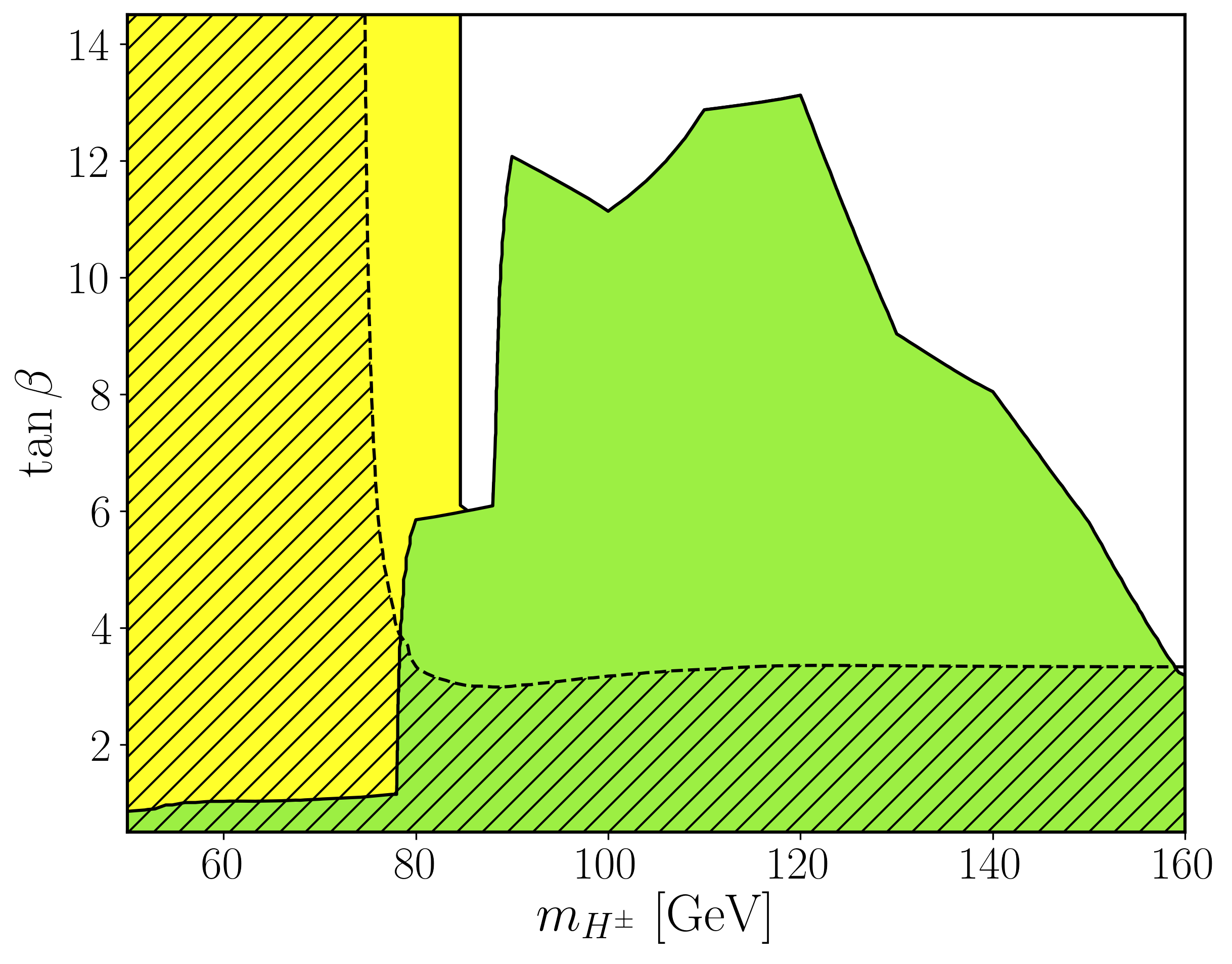}
	\end{minipage}
\caption{Constraints on the  $h\equiv\text{SM}$ scenario Type-III (left) and Type-I (right) from Higgs searches at the (LHC \& LEP), in the $(m_{H^{\pm}}, \tan\beta)$ plane. The hatched area is excluded by a mismatch between the properties of $h$ and those of the observed Higgs boson, and the (green and yellow) areas are excluded by the searches for additional Higgs bosons.}\label{fig1}
\end{figure}	
We now examine the light charged Higgs decays in the framework of 2HDM type-III. The relaxing bounds on charged Higgs mass in this  model is, in fact, due to the contribution of the non diagonal Yukawa couplings in the amplitude of the B meson decay, $\overline{B}\to X_s\gamma$.\\
Figure~\ref{fig2} shows that the decay width of charged Higgs is not dominated by $\hp \ra \tau \nu $ and $\hp \ra t^{*} \overline{b} $ decays. The decay channel $\hp \ra \mu \nu$ becomes the dominant decay mode over almost the whole allowed parameter space region, with  a branching ratio of 80\%.  To understand the reason behind the dominance of $ \hp \ra \mu $ over the other fermionic channels $ t^{*} \overline{b} $, $ c \overline {s} $ and $ \tau\nu$, we will study the effect of the contribution of the free parameters derived from the non diagonal elements of the Yukawa matrix on the charged Higgs width, we recall that the partial decay width of the Charged Higgs into two leptons is as follows:
\begin{eqnarray}
	\Gamma(H^+ \ra l^+_j \nu_i) = \frac{G_\mu \mhp}{4 \sqrt{2} \pi} m_{l_j}^2 
	\left((\xi^\ell_A)_{ij}\right)^2 \left(1- \frac{m_l^2}{\mhp^2}\right)^3
\end{eqnarray}
where $i = j$, $m_l \ll \mhp$ and $(\xi^\ell_A)_{ij} = \tan \beta - \frac{1}{\sqrt{2} \cos \beta} \chi_{ij}^l$. 
\begin{figure}[H]	
	\begin{minipage}{0.52\textwidth}
		\centering
		\includegraphics[height=5.6cm,width=7.6cm]{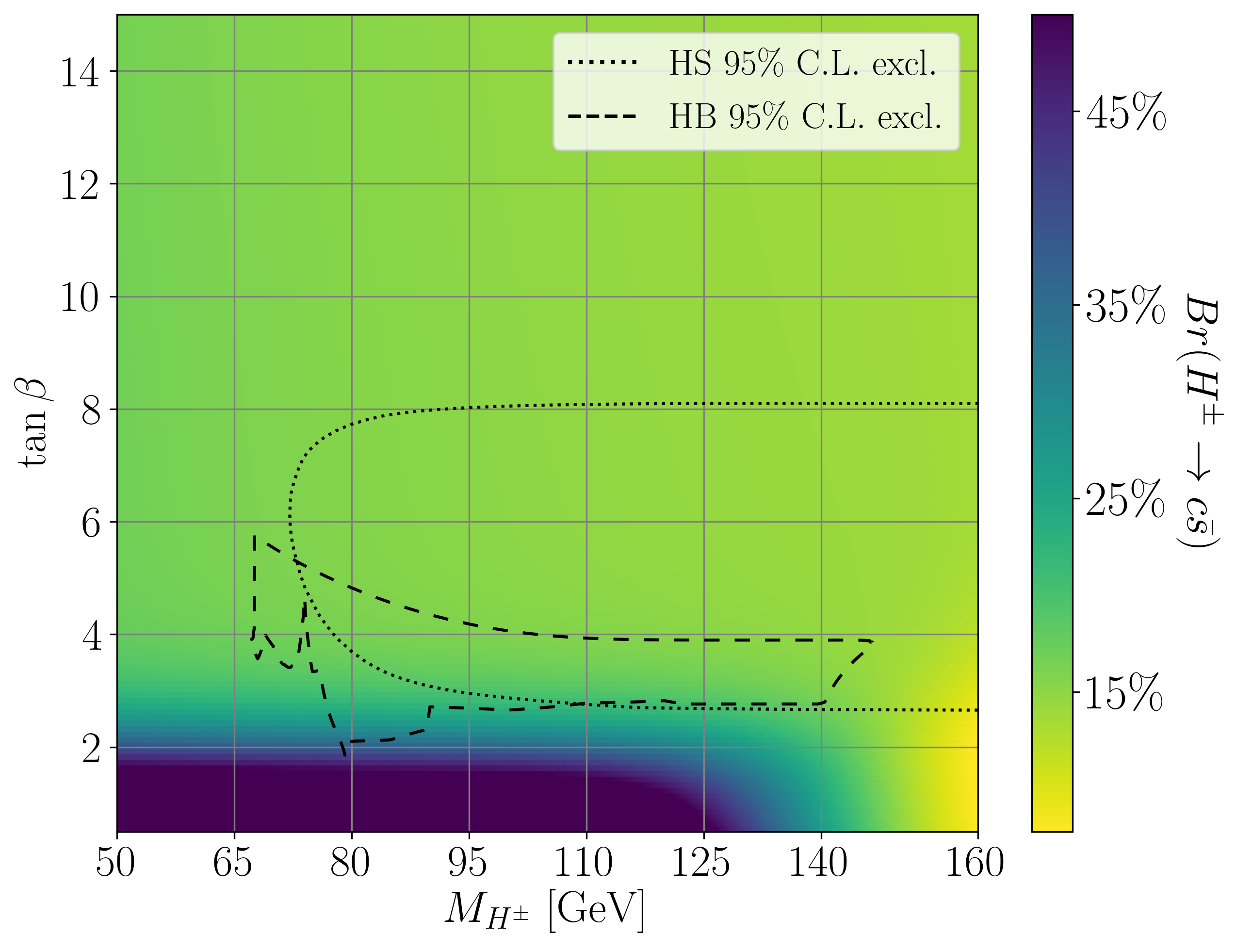}
	\end{minipage}
	\begin{minipage}{0.52\textwidth}
		\centering
		\includegraphics[height=5.6cm,width=7.6cm]{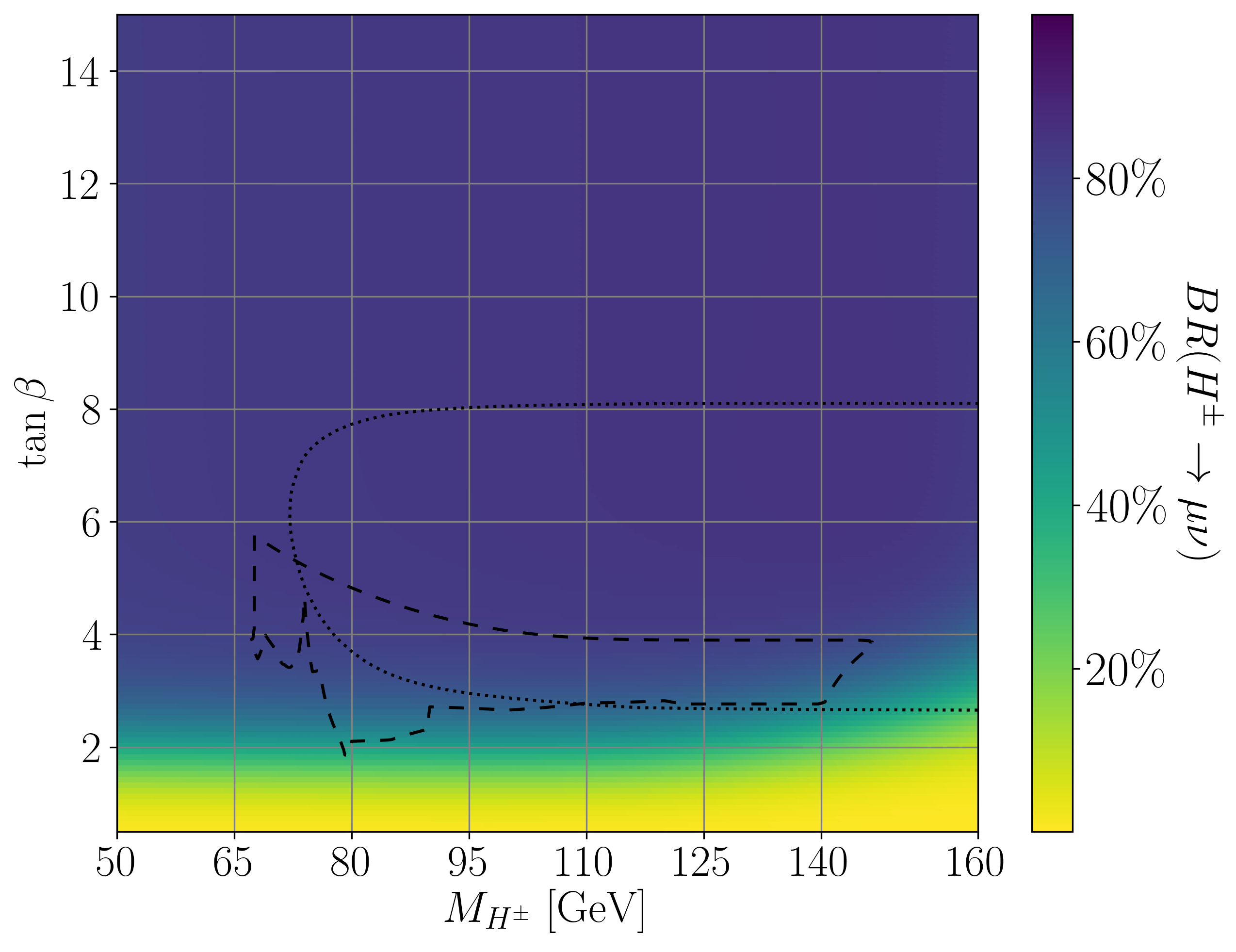}
	\end{minipage}
	\begin{minipage}{0.52\textwidth}
		\centering
		\includegraphics[height=5.6cm,width=7.6cm]{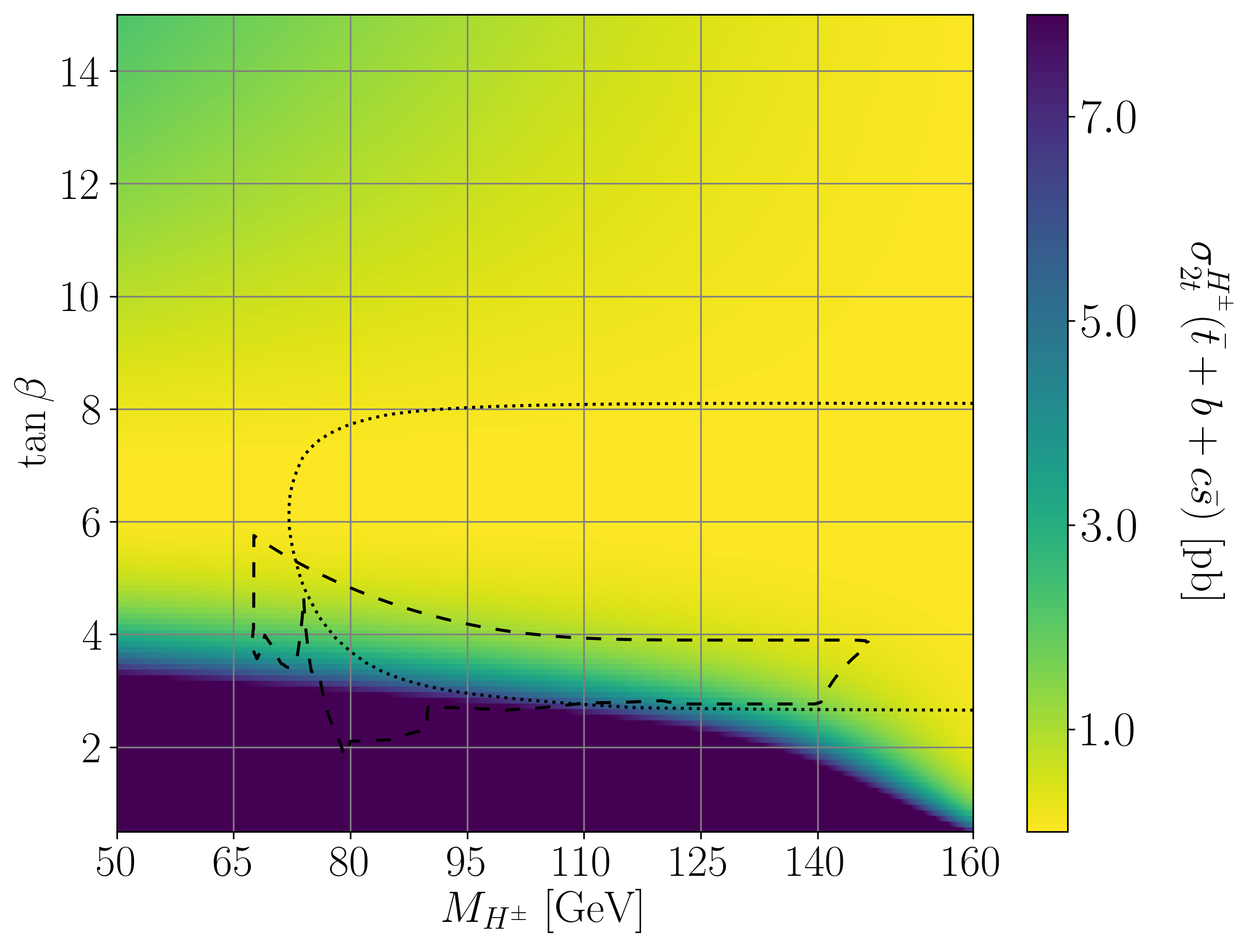}
	\end{minipage}
	\begin{minipage}{0.52\textwidth}
		\centering
		\includegraphics[height=5.6cm,width=7.6cm]{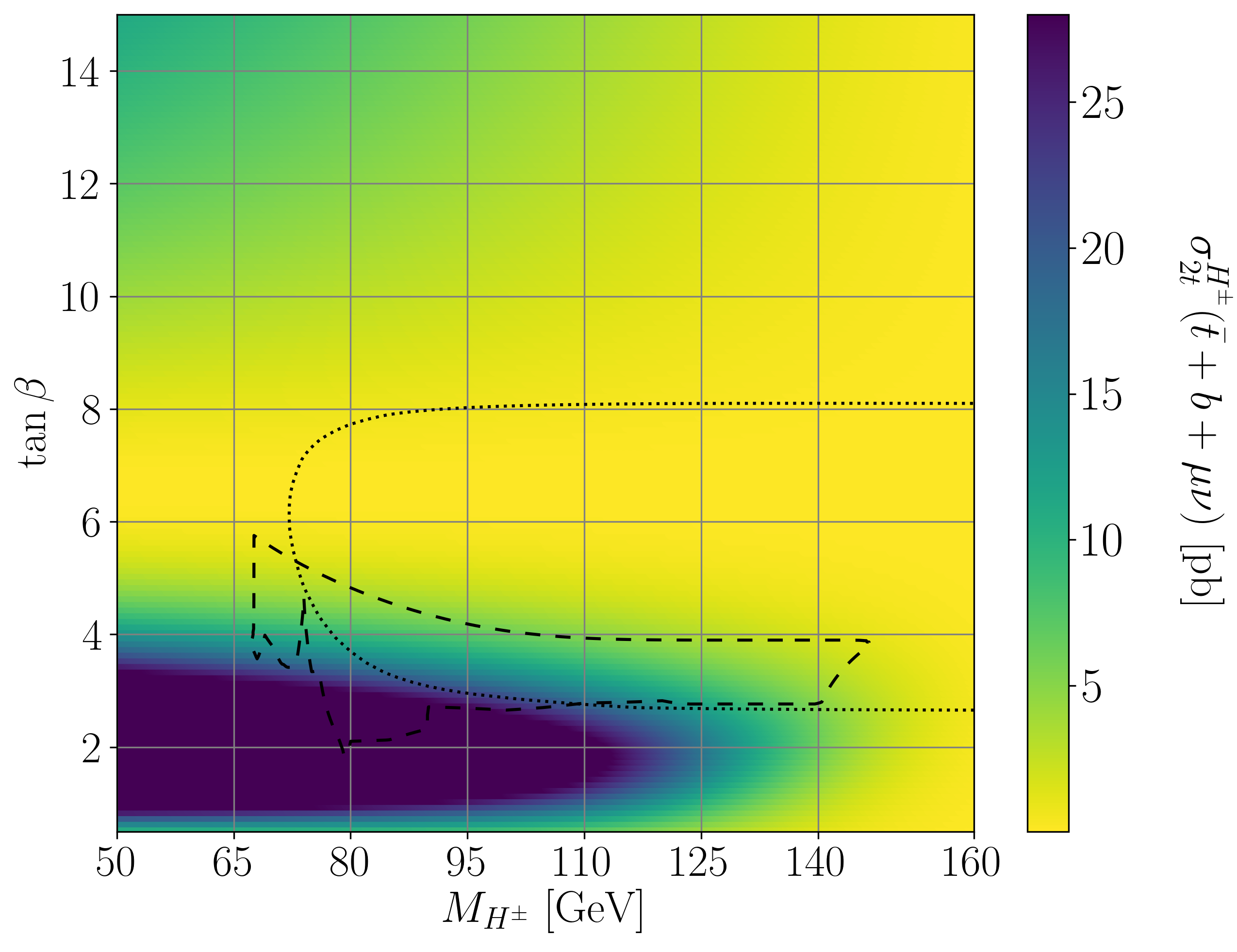}
	\end{minipage}	
\caption{The $BR(H^{\pm}\rightarrow XY )$ (up) and ${\sigma^{H^{\pm}}_{2t}(\bar{t}+b+XY)}$ (down) mapped over the $(m_{H^{\pm}} , \tan\beta)$ plane. For $XY \equiv c\bar{s}$ (down left) and  $XY \equiv \mu\nu $ (right). In each plot, the boundaries of the (green \& yellow) and the hatched exclusion regions of Figure \ref{fig1} are also shown as a dashed and a dotted black line, respectively.}\label{fig2}
\end{figure}
Note that $|\chi_{22}^l| >|\chi_{11,33}^l|$, therefore the term $m_\mu^2 (\tan \beta - \frac{1}{\sqrt{2} \cos \beta} \chi_{22}^l)$ is dominant over the other couplings when the channel $\hp \ra t^{*}b$ is closed, especially $\tau \nu$. Furthermore, one can read from figure 2 that Type-III could predict large production rate of $\mu\nu$ and $c\bar{s}$ from charged Higgs decay, where \textcolor{black}{${\sigma^{H^{\pm}}_{2t}(\bar{t}+b+\mu\nu)}$ and ${\sigma^{H^{\pm}}_{2t}(\bar{t}+b+c\bar{s})}$} could reach values of \textcolor{black}{23 pb and 6 pb} respectively in the range of masses varying from 72 GeV to 150 GeV, due to the higher branching ratios of charged Higgs decays and its production rate from $pp\to t\bar{t}$ process.
\begin{figure}[H]	
	\centering
	\includegraphics[height=12.cm,width=15.5cm]{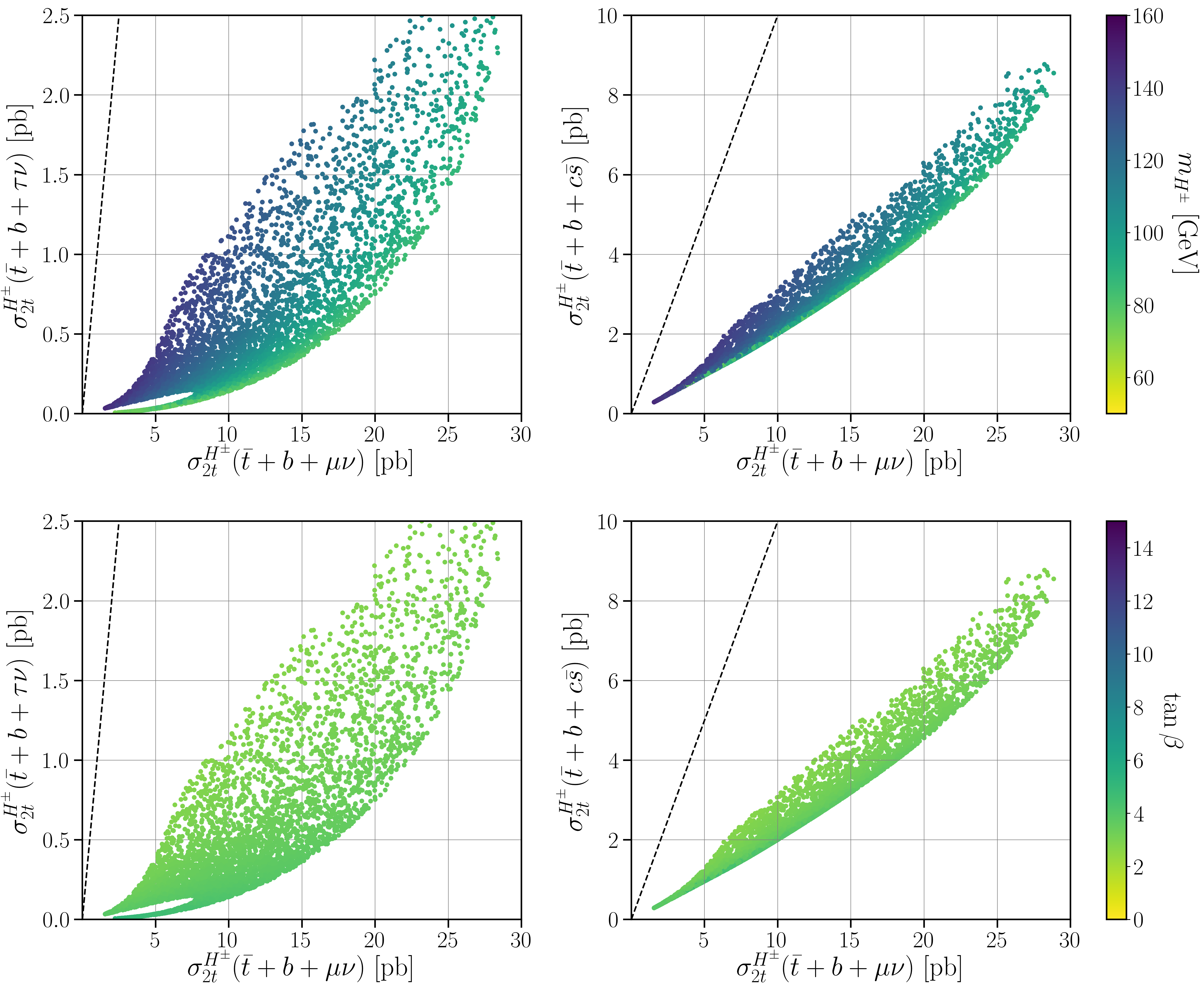}	
	\caption{${\sigma^{H^{\pm}}_{2t}(\bar{t}+b+\mu\nu)}$ v.s ${\sigma^{H^{\pm}}_{2t}(\bar{t}+b+XY)}$ with XY$\equiv \tau\nu$ (left) and XY $\equiv c\bar{s}$ (right). The color bar shows the mass of the charged Higgs boson (up) and $\tan\beta$ (down).}\label{fig44}
\end{figure}
In Figure \ref{fig44}, we show the correlation between production and decay of light charged Higgs into $\mu\nu$ and $\tau\nu$ (upper-left) and $cs$ (upper-right) mapping on charged Higgs mass $m_{H^\pm}$ (upper panels) and $\tan\beta$ (lower panels). It is clear from the plots that,  ${\sigma^{H^{\pm}}_{2t}(\bar{t}+b+\mu\nu)}$ is about one order of magnitude larger for intermediate values of $\tan\beta$ and within 80 GeV to 120 GeV for charged Higgs mass. Clearly, we have checked that $BR(t\to H^\pm b)\times BR(H^\pm \to \tau\nu)$ is below the current limit from CMS and ATLAS and below  0.01\%.
	\subsection{Part-II: $H$ Scenario}
In this scenario, we assume that the Higgs-like particle is $H$ with $m_H=125$ GeV, while $m_h$ is fixed at 95 GeV. We perform a systematic scan over the 2HDM parameters as indicated in Table \ref{2hdm_par}.
Figure \ref{fig4p} illustrates the allowed parameter space after taking into account most update  theoretical and experimental constraints. In this configuration, a large region of 2HDM parameter space as well as large $\tan\beta$ are allowed, compared to the first configuration (standard hierarchy). In order to understand the allowed region, we plotted various contributions arising from different processes (see Figure  \ref{fig12} (left) In Appendix) due to parameter relations to constraint ($m_{H^\pm}, \tan\beta$) space. Most allowed region in Figure \ref{fig4p}(left) comes from $pp\to gg \to A \to \mu^+ \mu^-$ channel due to Yukawa coupling in 2HDM type-III framework. Note that most of scanned region is compatible with the observed Higgs boson state at 95$\%$ C.L.  Negative search comes from $pp \to tbH^\pm$ followed by $H^\pm \to \tau \nu$ at the LHC in 36.1 $fb^{-1}$ of 
data at $\sqrt{s} = 13$ TeV. Such limit exclude large parameter space in 2HDM type-I.  As can be seen from table \ref{constraints_percent}, the number of the allowed points, by the theoretical and experimental constraints, is significant in the inverted hierarchy scenario, where $45\%$ of $10^6$ points are allowed after scanning over 2HDM parameter space.
After analyzing the allowed parameter space regions, we study the charged Higgs decay in such scenario.
\begin{figure}[t]
	\centering
	\begin{minipage}{0.51\textwidth}
		\centering
		\includegraphics[height=6.cm,width=7.0cm]{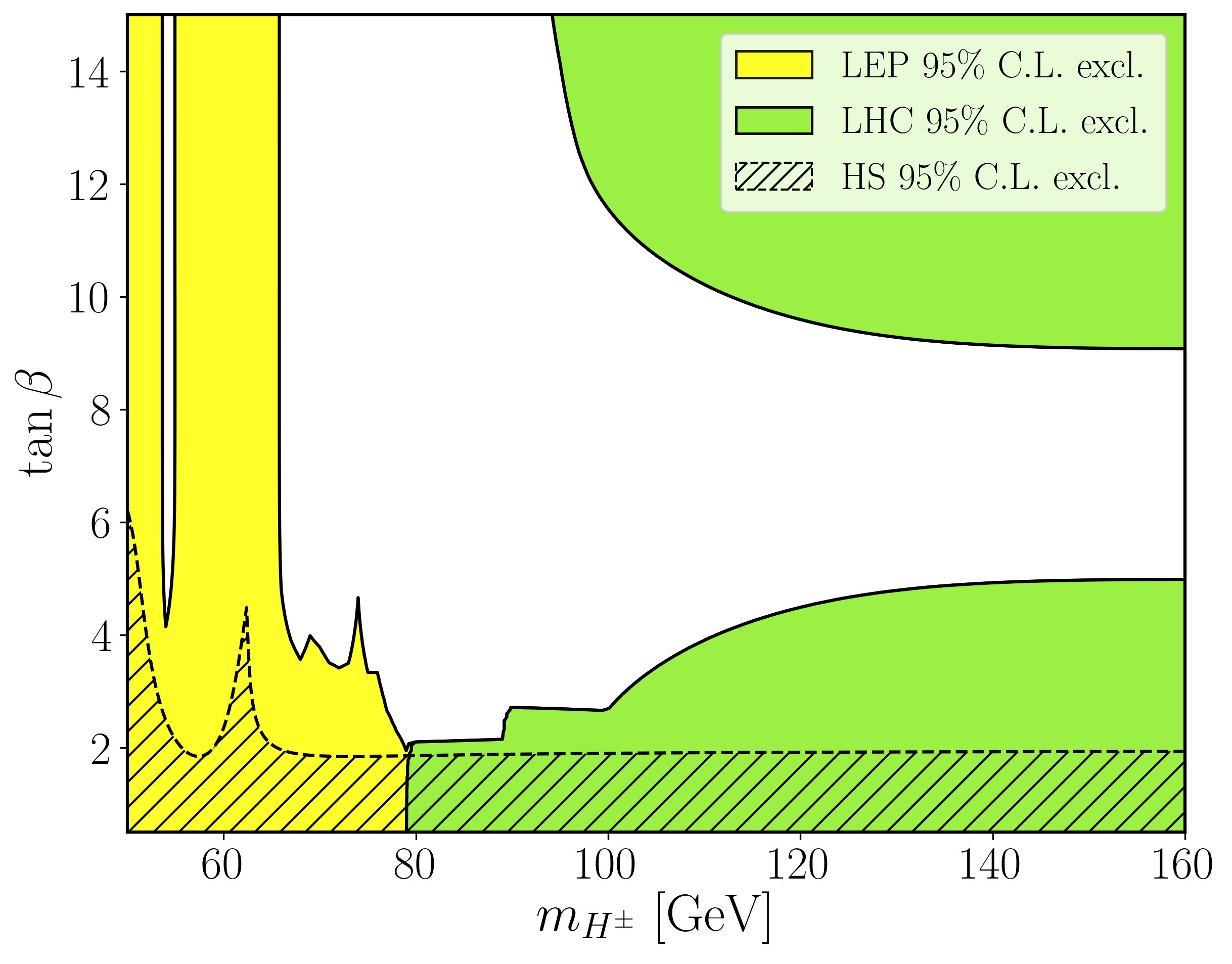}
	\end{minipage}
	\begin{minipage}{0.45\textwidth}
		\centering
		\includegraphics[height=6.cm,width=7.0cm]{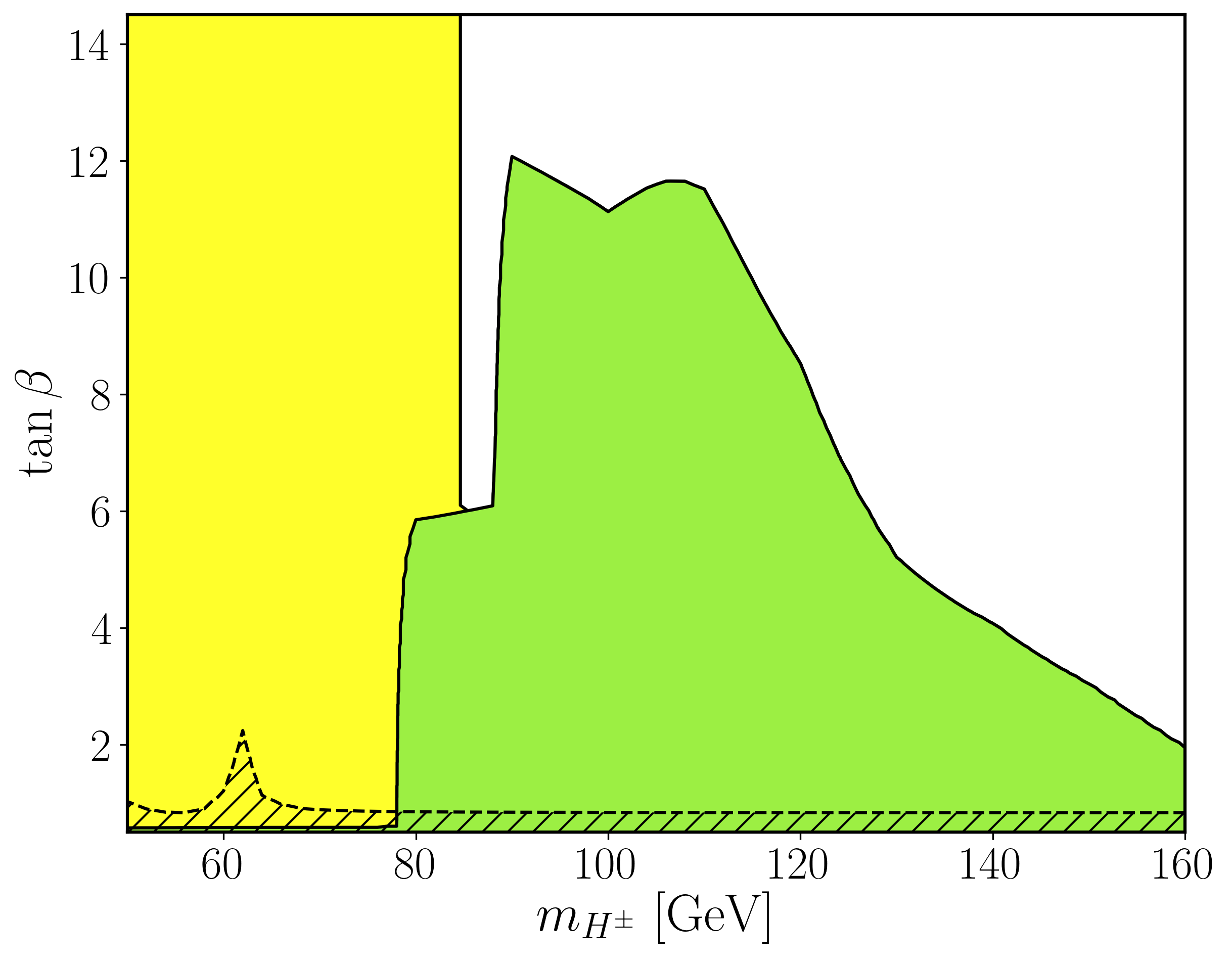}
	\end{minipage}
\caption{Constraints on the  $H\equiv\text{SM}$ scenario  Type-III (left) and Type-I (right) from Higgs searches at the (LHC \& LEP), in the $(m_{H^{\pm}}, \tan\beta)$ plane. The hatched area is excluded by a mismatch between the properties of $h$ and those of the observed Higgs boson, and the (green and yellow) areas are excluded by the searches for additional Higgs bosons.}\label{fig4p}.
\end{figure}	
\begin{figure}[h]	
	\begin{minipage}{0.52\textwidth}
		\centering
		\includegraphics[height=5.6cm,width=7.0cm]{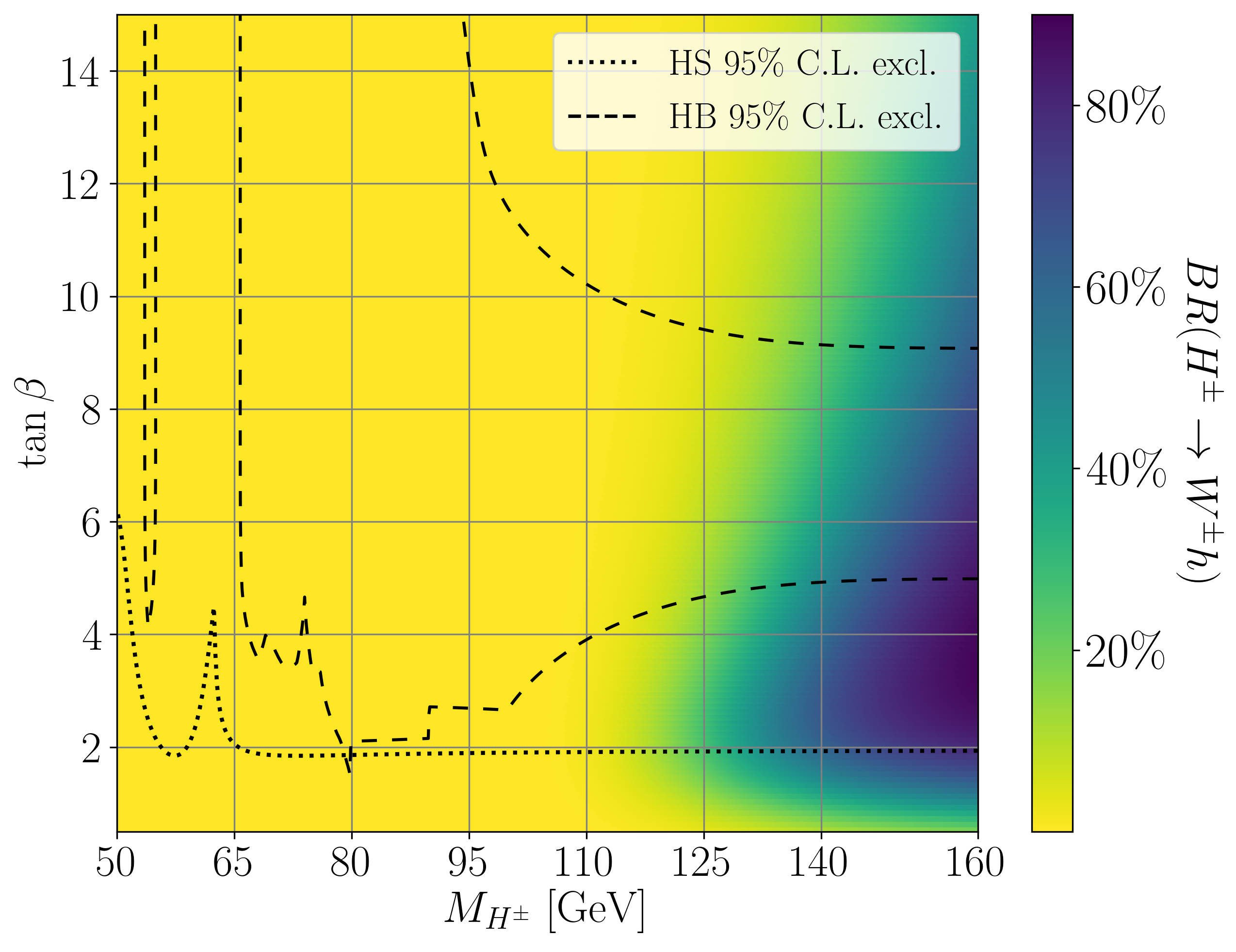}
	\end{minipage}
	\begin{minipage}{0.52\textwidth}
		\centering
		\includegraphics[height=5.6cm,width=7.0cm]{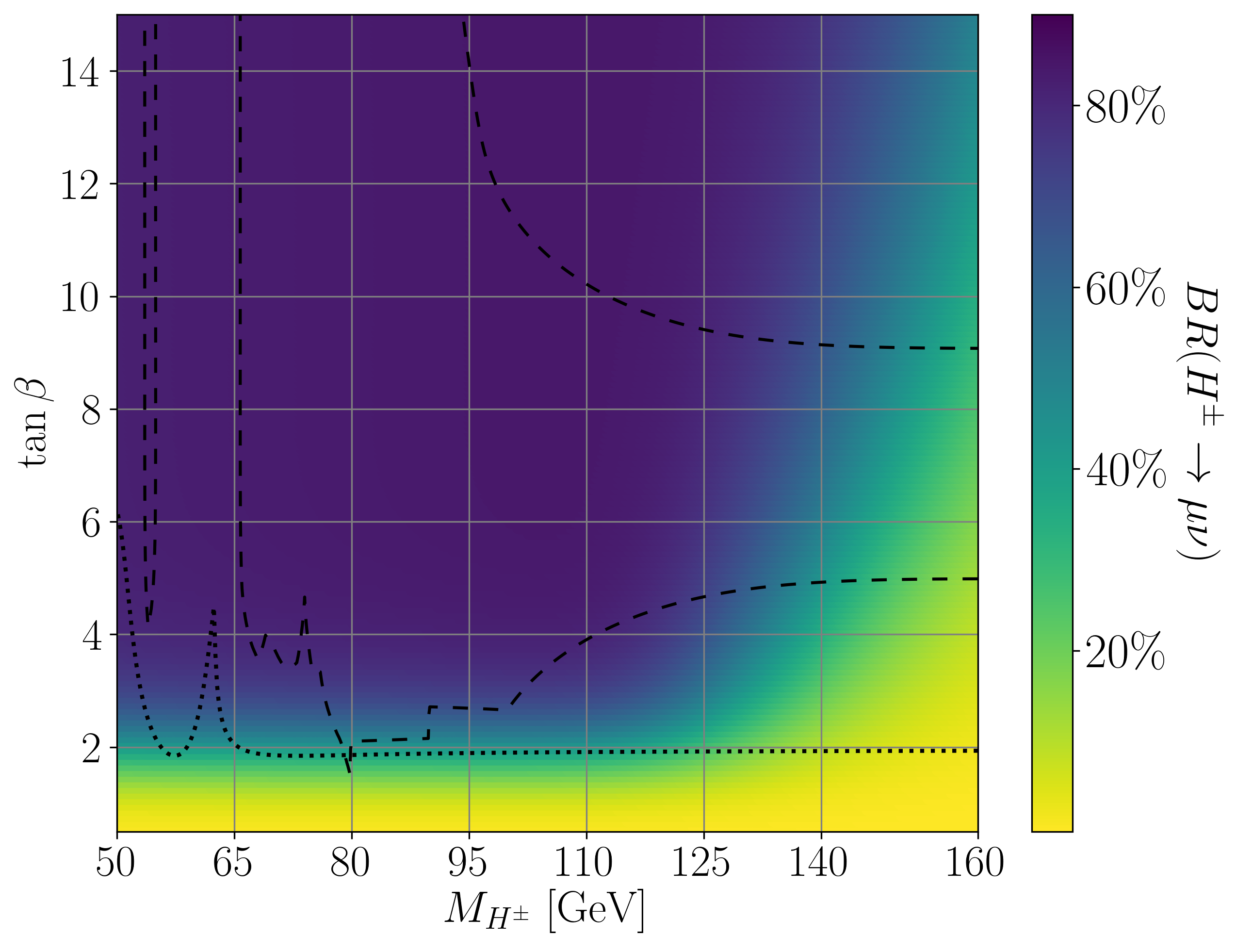}
	\end{minipage}
	\begin{minipage}{0.52\textwidth}
		\centering
		\includegraphics[height=5.6cm,width=7.0cm]{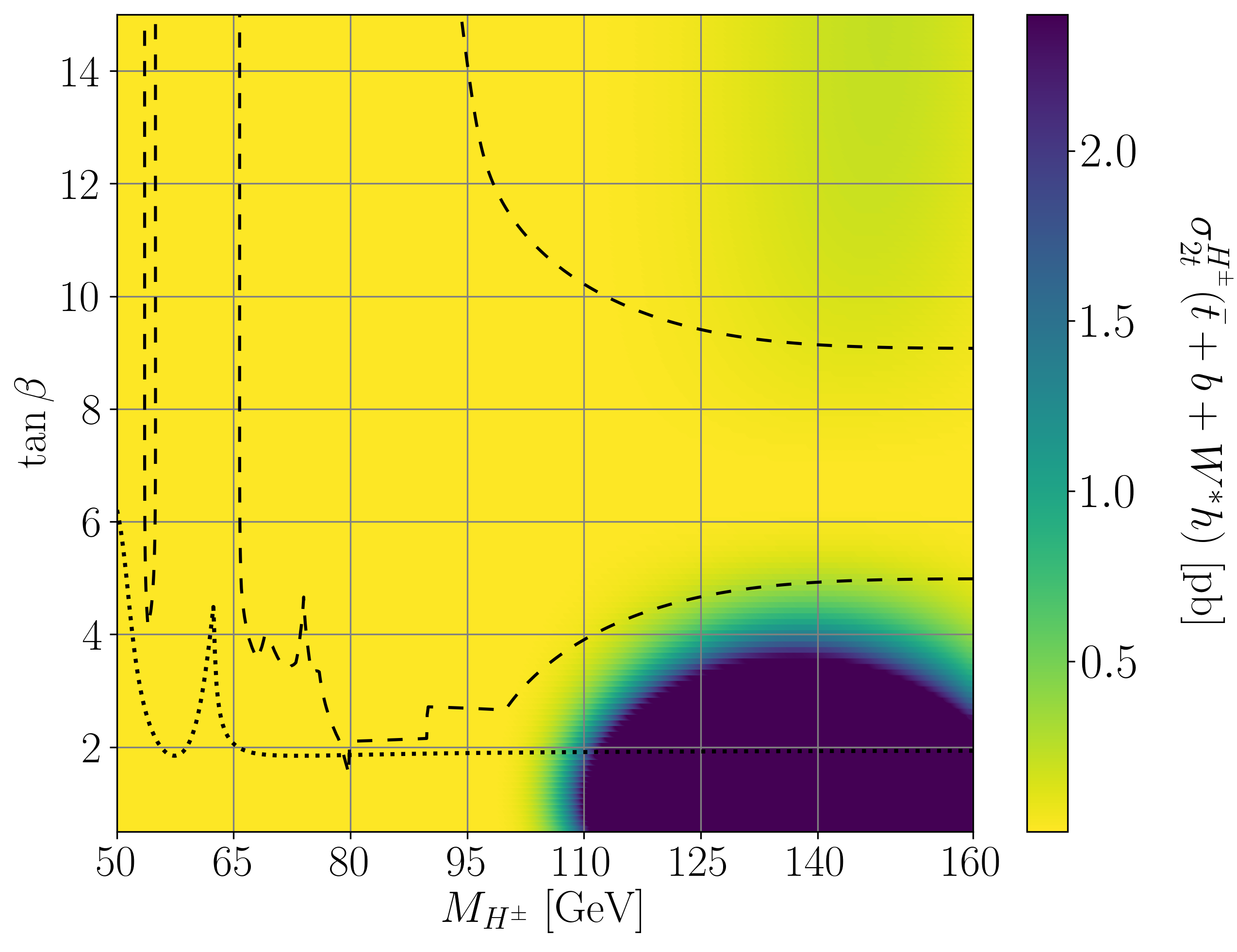}
	\end{minipage}
	\begin{minipage}{0.52\textwidth}
		\centering
		\includegraphics[height=5.6cm,width=7.0cm]{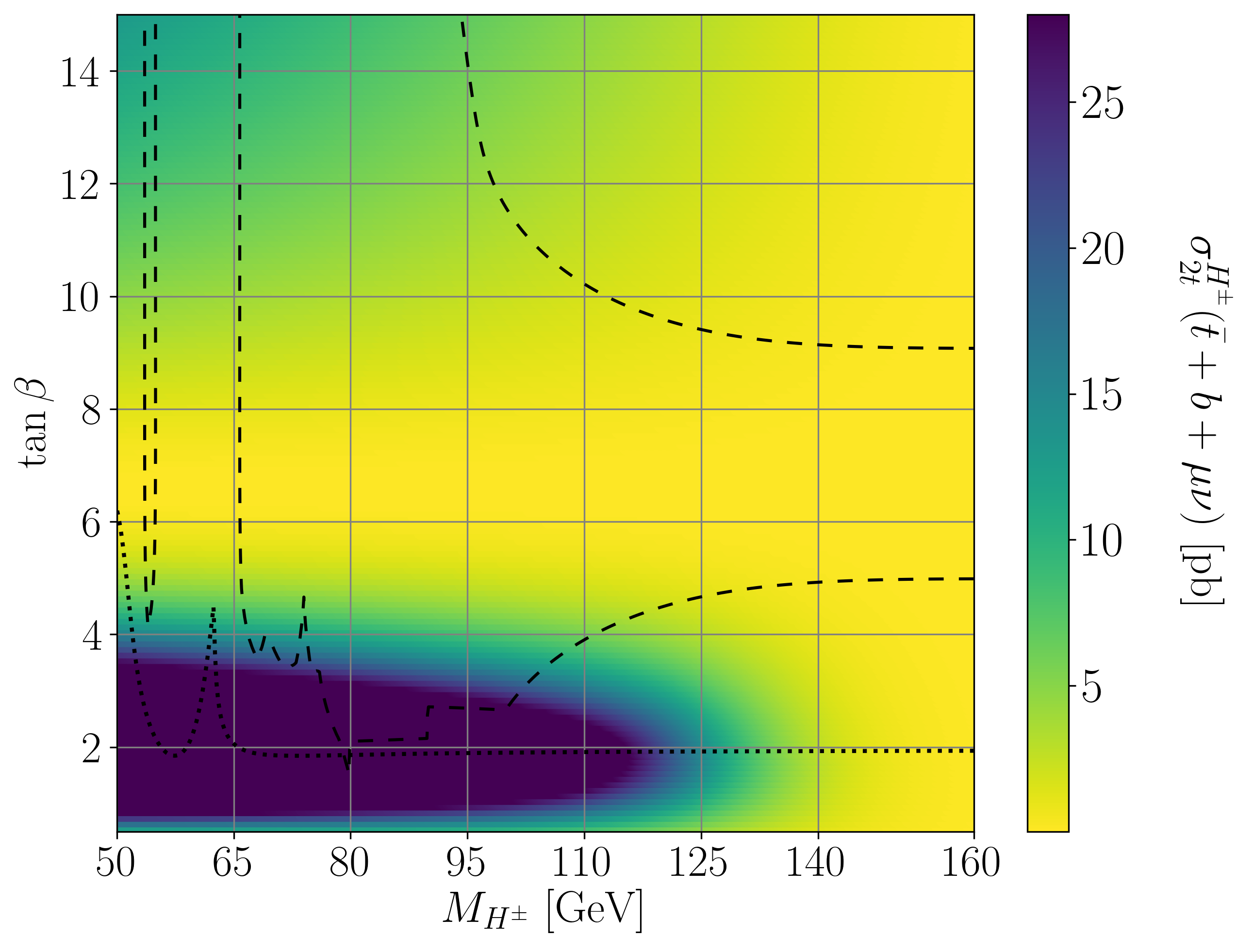}
	\end{minipage}	
	\caption{The $BR(H^{\pm}\rightarrow XY )$ (up) and ${\sigma^{H^{\pm}}_{2t}(\bar{t}+b+XY)}$ (down)
		mapped over the $(m_{H^{\pm}} , \tan\beta)$ plane. For $XY \equiv W^(*)h$ (left) and  $XY \equiv \mu\nu $ (right). In
		each plot, the boundaries of the (green \& yellow) and the hatched exclusion regions of Fig \ref{fig4p} are also shown as a dashed and a dotted black line, respectively.}\label{fig5}
\end{figure}

\begin{figure}[h]	
	\centering
	\includegraphics[height=12.5cm,width=15.5cm]{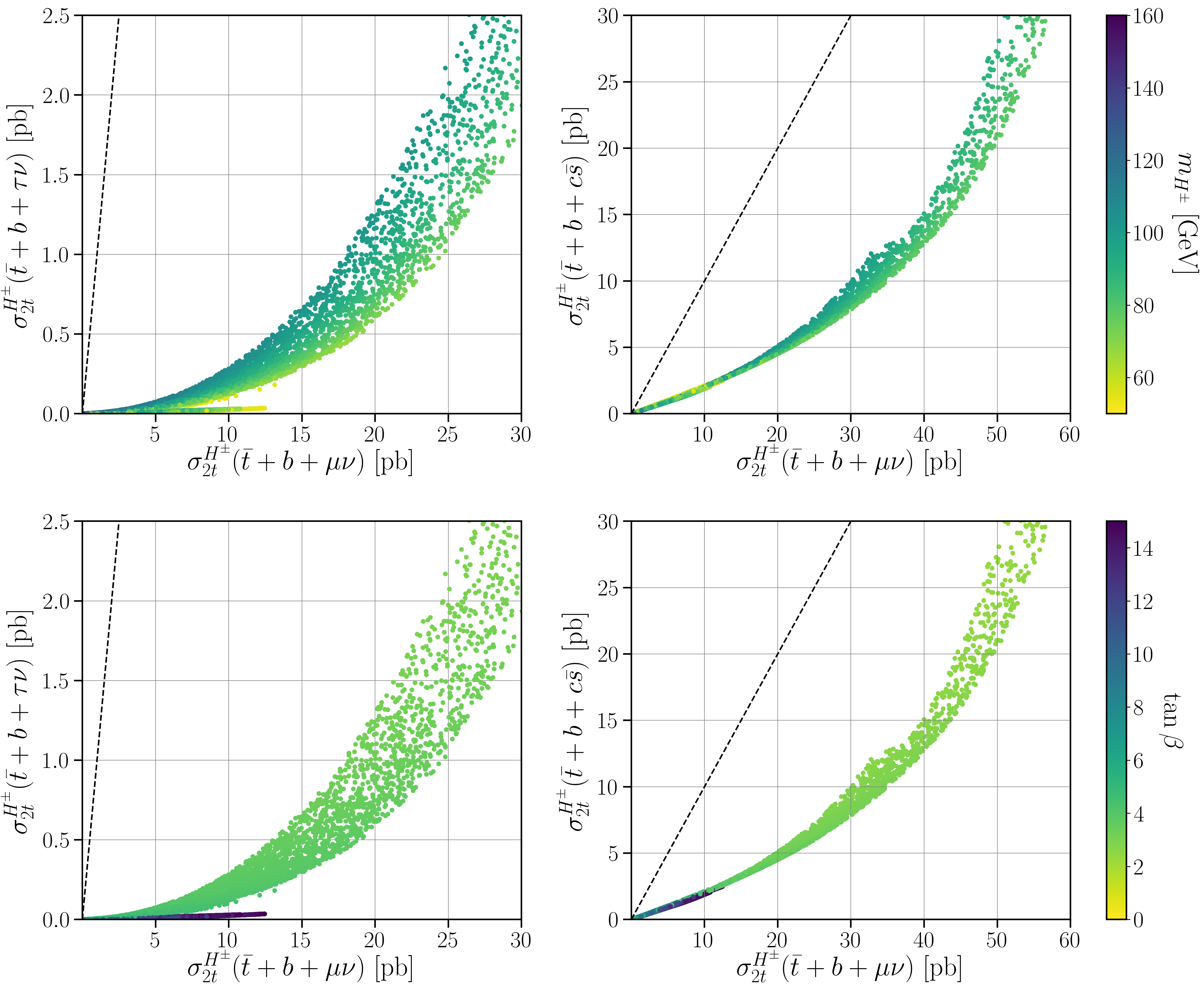}	
	\caption{${\sigma^{H^{\pm}}_{2t}(\bar{t}+b+\mu\nu)}$ v.s ${\sigma^{H^{\pm}}_{2t}(\bar{t}+b+XY)}$ with XY$\equiv \tau\nu$ (left) and XY $\equiv c\bar{s}$ (right). The color bar shows the mass of the charged Higgs boson (up) and $\tan\beta$ (down).}\label{fig6}
\end{figure}
\begin{table}[H]
	\centering
	\renewcommand{\arraystretch}{1} %
	\setlength{\tabcolsep}{6pt}
	\begin{tabular}{|c|c|c|c|c|} \hline
		&\makecell{Allowed by theoretical \\ constraints} & \makecell{Allowed by \\ HiggsBounds} & \makecell{Allowed by \\ HiggsSignals} & \makecell{Allowed \\ by all constraints}  \\ \hline \hline
		$h$ Scenario	&$62.33\%$  & $7.08\%$  &  $26.4\%$ & $4.31\%$ \\ \hline 
		$H$ Scenario	&$100\%$  & $45.08\%$  & $90.2\%$  &  $45.07\%$ \\ \hline
	\end{tabular}
	\caption{Details of the allowed points in parameter space after imposing theoretical and experimental constraints.} \label{constraints_percent}
\end{table} 

Figure~\ref{fig5} shows the branching ratios for the main decays of charged Higgs in 2HDM type-III. In such configuration, the $ H^\pm \ra W^{(*)} h $ channel could be open and thus it could strongly compete with the other channels like $ \tau \nu $, $ cs $ and $ t^*b $, because of the enhancement of its coupling $ h \hp W^\mp $ in the alignment limit.
The charged Higgs decay width is dominated by the two channels $ \hp \ra W^{(*)} h $ and $ \hp \ra \mu \nu $ over the other fermionic channels. The former can reach a maximum width around 80\% when the vector boson is being off shell, therefore the $ H^\pm \ra W^{(*)} h$ channel would be open, whereas $\hp \ra \mu \nu$ becomes significant due to the contribution of the off diagonal elements in the Yukawa couplings. 2HDM type-III predicts then a scenario where \textcolor{black}{${\sigma^{H^{\pm}}_{2t}(\bar{t}+b+\mu\nu)}$}  becomes significant and yields \textcolor{black}{27 pb} at 13 TeV for light $H^\pm$ and intermidiate  $\tan \beta$. On the other hand, \textcolor{black}{${\sigma^{H^{\pm}}_{2t}(\bar{t}+b+W^{(*)}h}$} is suppressed by the small production rate of charged Higgs via top decay for large $\tan \beta$.

As in Figure~\ref{fig6} (similar to Figure~\ref{fig44}), we present the corresponding correlation from ${\sigma^{H^{\pm}}_{2t}(\bar{t}+b+\mu\nu)}$  with ${\sigma^{H^{\pm}}_{2t}(\bar{t}+b+c\bar{s})}$  and ${\sigma^{H^{\pm}}_{2t}(\bar{t}+b+\tau\nu)}$  mapping on charged Higgs mass (upper panels) and on $\tan\beta$ (lower panels). It is clear that ${\sigma^{H^{\pm}}_{2t}(\bar{t}+b+\mu\nu)}$  can one order of magnitude over $\tau\nu$ and $cs$ modes for roughly $m_{H^\pm}$ in 80 GeV $\sim$120 GeV range and $\tan\beta$ in $2\sim5$ range.
 It is therefore clear that  ${\sigma^{H^{\pm}}_{2t}(\bar{t}+b+\mu\nu)}$  can provide a potential alternative discovery channel for light charged Higgs boson at the LHC in the 2HDM-type-III.

\subsection{Benchmark points}
Table \ref{tab8} summarizes some Benchmark Points (BPs) for each scenario. All these BPs satisfy the most update theoretical and experimental constraints. For every BP in the table, we give the cross sections of $t+b+XY$, where $XY=cs, t^* b, W^(*)h, \tau\nu$ and $\mu\nu$ signatures. Furthermore, the production rate of $t+b+\mu\nu$ can reach few pico barn in scenario where $H$ is the SM state. Hence, in the standard hierarchy ($h$ is assumed SM-like), the aforementioned cross sections are still sizable. 


\begin{table}[t]
	\centering
	\renewcommand{\arraystretch}{1} %
	\setlength{\tabcolsep}{16pt}
 \begin{tabular}{|c|c|c|c|c|c|}
	\hline
	\multicolumn{6}{|c|}{Selected BPs in Part I} \\\hline
	Parameters & BP1 & BP2   & BP3 & BP4 & BP5\\\hline
	\multicolumn{6}{|c|}{The Higgs masses are in GeV} \\\hline
	$m_h$ &  $125$ & $125$ & $125$ & $125$& $125$\\
	$m_H$ &  $135$ & $135$  & $135$ & $135$& $135$\\
	$m_A$ &  $220$ & $220$  & $220$& $220$ & $220$\\  
	$m_{H^\pm}$ &  $110$ & $93.2$  & $99.4$& $91.4$& $102$\\ 
	$\cos{(\beta-\alpha)}$ & $0.145$ & $0.145$ & $0.145$& $0.145$& $0.145$\\ 
	$\tan\beta$ & $2.9$ & $3.2$ & $3.7$& $4.3$& $3$ \\
	$m_{12}^2$ & $4815.35$ & $4448.39$ & $3935.5$& $3447.28$ & $4687.5$\\ \hline\hline
	\multicolumn{6}{|c|}{${\sigma^{H^{\pm}}_{2t}(\bar{t}+b+XY)}$ [pb]} \\\hline
	$c\bar{s}$&$6.2$ &$5.76$ & $2.48$ & $1.32$& $6.48$\\
	$t^*\bar{b}$&$0.2$& $-$ &$0.002$& $-$& $0.04$\\
	$W^(*)h$&$-$&$-$&$-$&$-$&$-$\\
	$\mu\nu$&$21.12$&$11.46$&$22.92$&$6.82$&$23.32$\\
	$\tau\nu$&$2$&$1.36$&$0.34$&$0.08$&$1.88$\\
	\hline\hline
	\multicolumn{6}{|c|}{Selected BPs in Part II} \\\hline
	Parameters & BP1 & BP2   & BP3 & BP4 & BP5\\\hline
	\multicolumn{6}{|c|}{The Higgs masses are in GeV} \\\hline
	$m_h$ &  $95$ & $95$ & $95$& $95$ & $95$ \\
	$m_H$ &  $125$ & $125$  & $125$& $125$  & $125$ \\
	$m_A$ &  $177$ & $177$  & $177$& $177$  & $177$ \\  
	$m_{H^\pm}$ &  $95$ & $90$  & $94$  & $99.6$  & $90.80$\\ 
	$\sin{(\beta-\alpha)}$ & $-0.05$ & $-0.05$ & $-0.05$&$-0.05$ & $-0.05$\\ 
	$\tan\beta$ & $3$ & $3.5$ & $4.2$   & $3.7$ & $3.2$ \\
	$m_{12}^2$ & $2707.5$ & $2383.96$ & $2033.53$     & $2273.14$ & $2569.39$\\ \hline\hline
	\multicolumn{6}{|c|}{${\sigma^{H^{\pm}}_{2t}(\bar{t}+b+XY)}$ [pb]} \\\hline
	$c\bar{s}$&$7.58$&$3.98$&$1.42$&$2.46$&$6.06$\\
	$t^*\bar{b}$&$-$&$-$&$-$&$0.003$&$-$\\
	$W^(*)h$&$-$&$-$&$-$&$0.00046$&$-$\\
	$\mu\nu$&$26.92$&$17.8$& $7.36$&$11.86$&$23.96$\\
	$\tau\nu$&$2.18$&$0.66$&$0.1$&$0.34$&$1.42$\\\hline	
 \end{tabular}	
\caption{Mass spectra, mixing angles, branching ratios and cross section (in pb) in each configuration} \label{tab8}
\end{table}


\section{Conclusions}

With the addition of a Higgs doublet to the SM, we can address some basic issues of the SM, for instance the question of the structure of EWSBs, the flavor enigma, the possibility of the existence of an additional CP violation, the mechanics of EW baryogenesis, the identification of appropriate targets for dark matter (DM), etc. This extra addition of scalar doublet provides a general 2HDM, which makes it possible to obtain a more accurate picture of the nature. Charged Higgs sector of general 2HDM has been thoroughly investigated in the existing literature as well as an interpretation projection from experiments collaborations. 

In this current work, we intended to explore the phenomenology of charged Higgs in 2HDM-type I and type-III, where the parameter region are similar, and both scenarios provide alternative ways to capture light charged Higgs. To be more accurate, the LHC collaborations are now in search of new states beyond the SM, such realization could be done in general 2HDM for example.  The main advantage of generating $pp\to t \bar t$  process is twofold. Firstly, the large cross-section for top anti-top are produced via QCD interactions which do not rely on the 2HDM parameters. Secondly, top quark may interact with charged Higgs boson as $t\to H^\pm b$ mode, where it is model dependent. Charged Higgs boson when produced, it may decay into standard particles. The so-called Yukawa couplings are related to the different manners by which the second Higgs doublet is realized  beyond SM.\\

In this work, we have specifically focused on the production of charged Higgs bosons via $pp\to t b H^\pm$ at the LHC with $\sqrt{s} = 13$ TeV in both 2HDM type-I and type-III. After considering all the up-to-date theoretical and experimental constraints, allowing $H^\pm \to \mu \nu$ instead of standard decay mode $H^\pm \to \tau \nu$, we have studied the final states $t+b+\mu\nu$ as potential discovery channel.  In doing so, we have compared the efficiency of that channels over a large range of charged Higgs mass, up to top quark mass when $\tan\beta$ is intermediate. The most interesting regions of the 2HDM to types are explored  to show significant potentially discoverable event rates. Therefore, with a view to letting the ATLAS and CMS collaborations either confirm or constraint this alternative, we eventually proposed a number of BPs in the following 2HDM type-I and -III final states that are suitable for further experimental investigation.

\newpage
\appendix
\section*{Appendix}
\begin{figure}[H]	
	\centering
	\includegraphics[height=6.8cm,width=15.7cm]{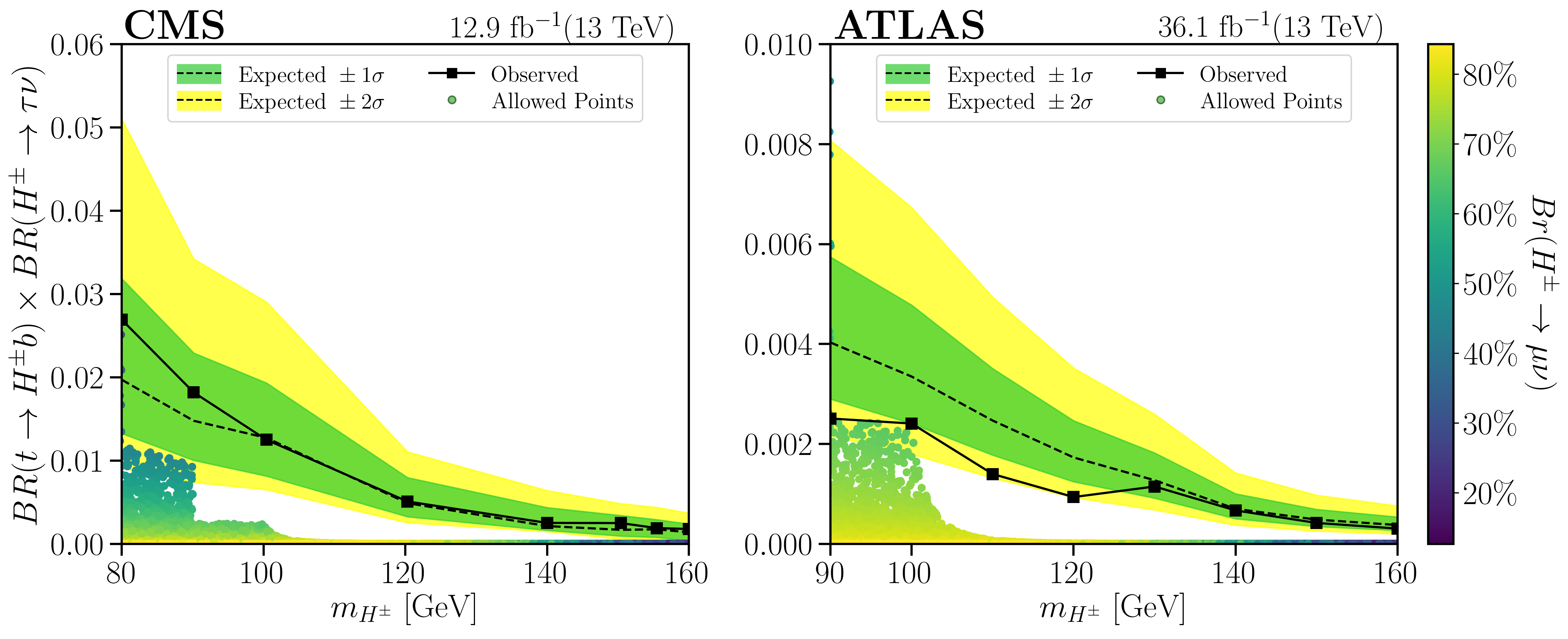}
\caption{Projection of our finding on ATLAS \cite{ATLAS:2018gfm} and CMS\cite{Sirunyan:2019hkq} results.}\label{fig10}
\end{figure}
\begin{figure}[H]
	\centering
	\includegraphics[height=8.cm,width=14.5cm]{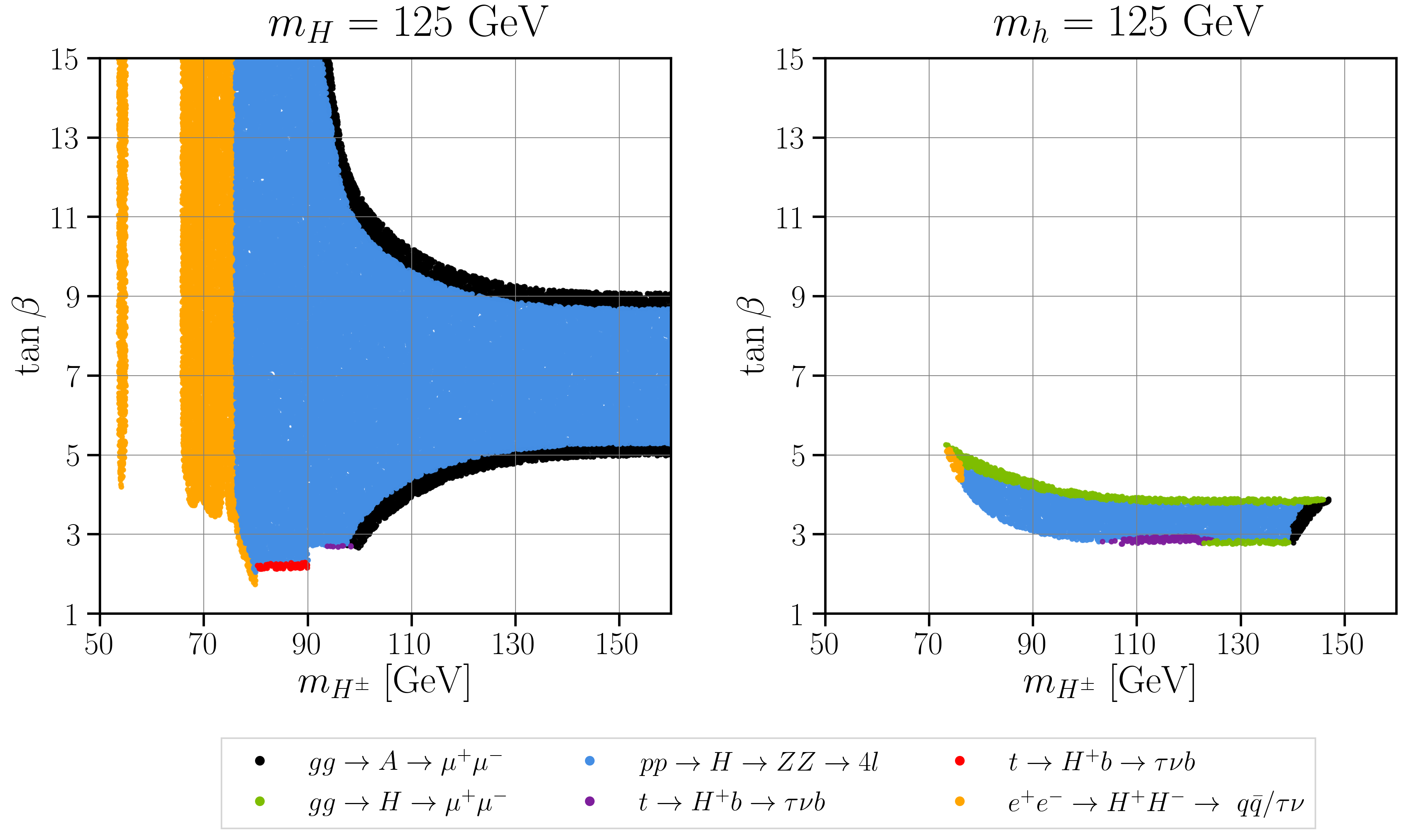}
\caption{The Allowed channels by all constraints for both configurations.}\label{fig12}	
\end{figure}

\end{document}